\documentclass[journal,10pt]{IEEEtran}

\usepackage{balance}
\usepackage[ruled]{algorithm2e}
\usepackage{amssymb}
\usepackage{mathbbol}
\usepackage{stmaryrd}
\usepackage{cite}
\usepackage{color}
\usepackage{multirow}
\usepackage{subfigure}
\usepackage{graphicx,times,amsmath}
\usepackage{epstopdf}
\usepackage{pifont}
\usepackage{url}
\usepackage{soul}

\soulregister\cite7
\soulregister\ref7

\begin{document}
\title{State-Aware Rate Adaptation for UAVs by Incorporating On-Board Sensors}

\author
{
		Shiyue~He,
		Wei~Wang,~\IEEEmembership{Member,~IEEE,}
		Hang~Yang, 
		Yang~Cao,~\IEEEmembership{Member,~IEEE,}
		Tao~Jiang,~\IEEEmembership{Fellow,~IEEE,}
		and~Qian~Zhang,~\IEEEmembership{Fellow,~IEEE}% <-this % stops a space

		\thanks{The work was supported in part by the National Key R\&D Program of China under Grant 2017YFE0121500, the National Natural Science Foundation of China (NSFC) with Grants 91738202, 61871441, 61601193, 61729101, 61631015, and 61720106001, the RGC under Contract CERG 16203719 and 16204418, the Fundamental Research Funds for the Central Universities with Grant number 2015ZDTD012, the Guangdong Natural Science Foundation No. 2017A030312008, Young Elite Scientists Sponsorship Program by CAST under Grant 2018QNRC001, the Key Laboratory of Dynamic Cognitive System of Electromagnetic Spectrum Space (Nanjing Univ. Aeronaut. Astronaut.), Ministry of Industry and Information Technology, Nanjing, 211106, China under Grant KF20181911.}

		% \thanks{This work was supported in part by National Science Foundation of China with Grant numbers 61729101, 61631015, and Fundamental Research Funds for the Central Universities with Grant number 2015ZDTD012}

		\thanks{S. He, W. Wang, H. Yang, Y. Cao, and T. Jiang are with the School of Electronic Information and Communications, Huazhong University of Science and Technology. 
		E-mail: \{shiyue\_he, weiwangw, hangyang, ycao, taojiang\}@hust.edu.cn.
		}
		\thanks{Q. Zhang is with the Department of Computer Science and Engineering, Hong Kong University of Science and Technology. E-mail: qianzh@cse.ust.hk.}
}

\maketitle
\begin{abstract}
Nowadays unmanned aerial vehicles (UAVs) are being widely applied to a wealth of civil and military applications. % MobiRate, abstract
Robust and high-throughput wireless communication is the crux of these UAV applications.
Yet, air-to-ground links suffer from time-varying channels induced by the agile mobility and dynamic environments. % softrate introduction
Rate adaptation algorithms are generally used to choose the optimal data rate based on the current channel conditions. %smartpilot, introduction 
State-of-the-art approaches leverage physical layer information for rate adaptation, and they work well under certain conditions. % smartpilot, abstract
However, the above protocols still have limitation under constantly changing flight states and environments for air-to-ground links. % smartpilot, abstract
To solve this problem, we propose StateRate, a state-optimized rate adaptation algorithm that fully exploits the characteristics of UAV systems using a hybrid deep learning model. 
The key observation is that the rate adaptation strategy needs to be adjusted according to motion-dependent channel models, which can be reflected by flight states.
In this work, the rate adaptation protocol is enhanced with the help of the on-board sensors in UAVs.  % MobiRate abstract 
To make full use of the sensor data, we introduce a learning-based prediction module by leveraging the internal state to dynamically store temporal features under variable flight states. 
We also present an online learning algorithm by employing the pre-trained model that adapts the rate adaptation algorithm to different environments. % MobiRate abstract
We implement our algorithm on a commercial UAV platform and evaluate it in various environments. 
The results demonstrate that our system outperforms the best-known rate adaptation algorithm up to 53\% in terms of throughput when the velocity is 2-6~$m/s$.

\end{abstract}


\section{Introduction}
\label{sec:intro}

Recent years have witnessed the rapid proliferation of unmanned aerial vehicles (UAVs) in both commercial and government applications.
Goldman Sachs Research forecasts that UAV market prospects will reach \$100 billion between 2016 and 2020~\cite{goldmansachs}. % RFly, introduction
A large number of UAV applications, such as aerial photograph~\cite{skyeyes}, search and rescue (SAR) missions~\cite{now_or_later, Wireless_Surveillance}, and live video streaming~\cite{flyingeye}, rely on wireless technologies to achieve reliable communication between UAVs in the air and terminals on the ground.
For example, the SAR missions require that rescue information is transmitted to the server within a short time live video streaming allows high-definition cameras and microphones to film videos and stream them to users on the ground. 

Due to the fast movements of UAVs and rapidly varying environments~\cite{Aerial_Channel,Energy_Tradeoff}, existing rate adaptation algorithms can not meet such high-quality transmission requirements.
How to build a stable and robust communication link between the UAV and the receiver on the ground remains an open problem. % smartpilot, introduction
Traditionally, rate adaptation algorithms have been proposed to optimize communication quality in wireless networks by dynamically tuning the modulation and coding scheme (MCS) according to channel variances.
Prior works use either packet loss~\cite{minstrel, ath9k, samplerate, ha-rraa} or signal-to-noise rate (SNR)~\cite{cara, charm, esnr, fara, softrate, smartpilot} extracted from the received signal strength indicator (RSSI) and channel state information (CSI) to pick the optimal bit rate.  %(softrate, introduction)
Most of the loss-based rate adaptation algorithms are highly inaccurate when applied to the mobile scenario because the packet loss is calculated by hundreds of frames and they fail to consider the impact induced by motion. %strider, related work; mobirate system design overview
To overcome the drawback, some loss-based algorithms are augmented with additional information for mobile and vehicular networks~\cite{brave}.
% These previous works are still inappropriate for the air-to-ground channel due to the flexible flight states for UAV. 
In contrast, the SNR-based algorithms assume that the channel states is constant within the channel coherence time to predict the channel states. % Aerial channel prediction and user scheduling in mobile drone hotspots, channel prediction algorithm
Although the SNR-based methods are more adaptive when the channel changes, the major deficiency in designing such a rate adaptation algorithm is that these methods require fine-tuning to enhance performance across different mobility-dependent channel models and fail to generalize to scenarios for UAVs. % softrate, related work; pensieve, introduction

\begin{figure}[t]
	% \vspace{0.1cm}
	\centering
	\includegraphics[width=0.5\textwidth]{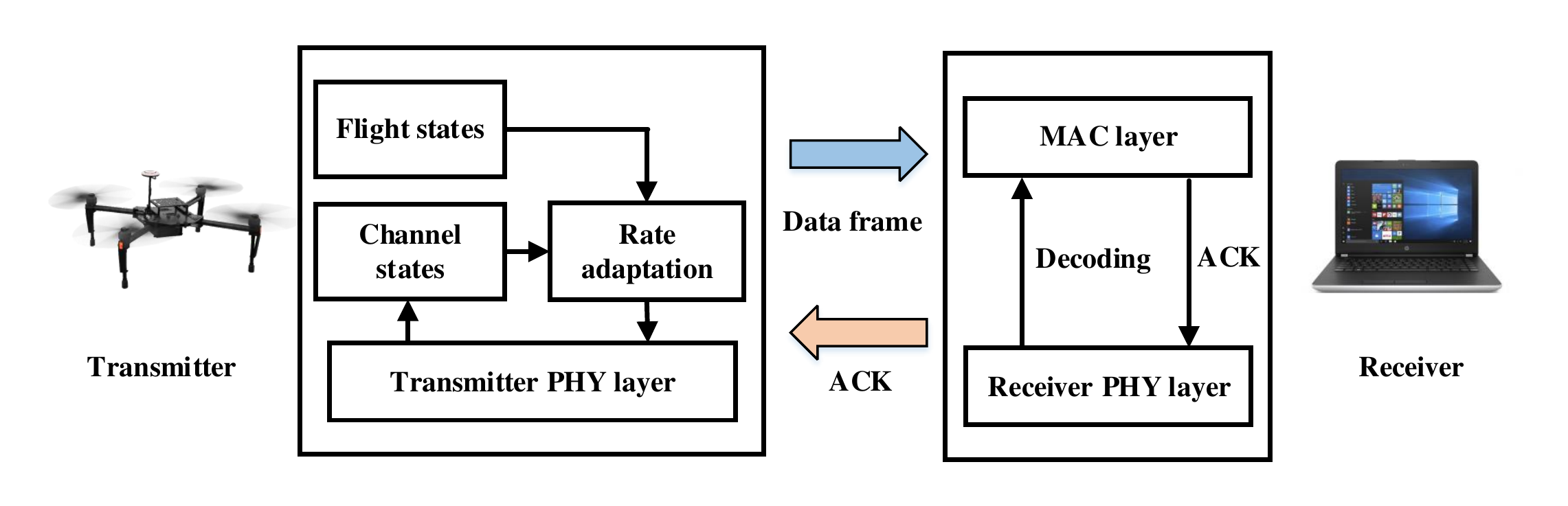}
	\caption{Rate adaptation overview.}
	\label{fig:schematic}
\end{figure}

\begin{figure*}
	\centering
	\begin{minipage}[b]{0.66\textwidth}
		\centering
		\subfigure[\scriptsize RSSI difference between adjacent packets]
		{
			\label{fig:snr_difference}
			\includegraphics[width=0.49\textwidth]{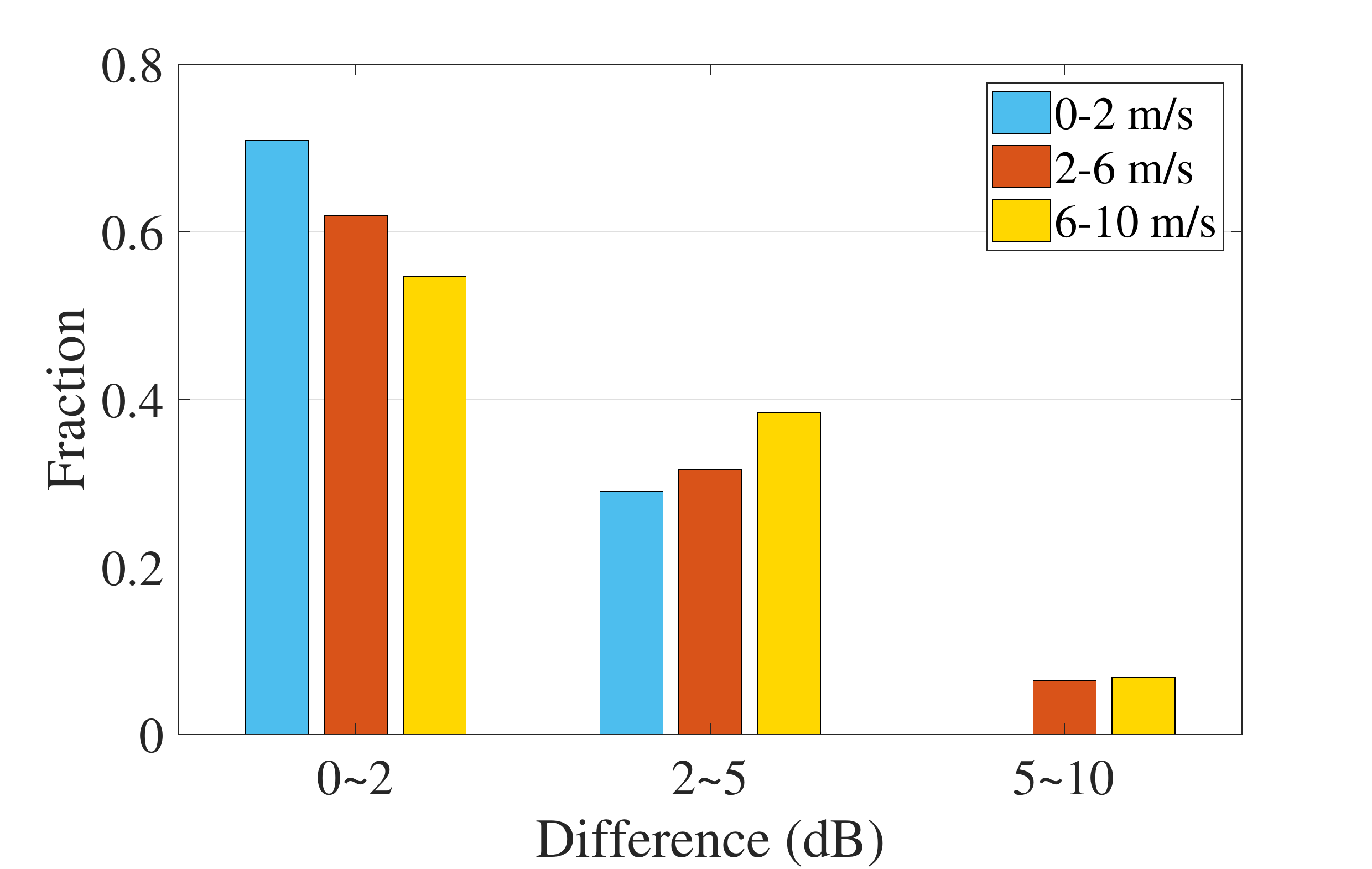}
		}
		\hspace{-0.7cm}
		\subfigure[\scriptsize CSI correlation]
		{
			\label{fig:csi_correlation}
			\includegraphics[width=0.49\textwidth]{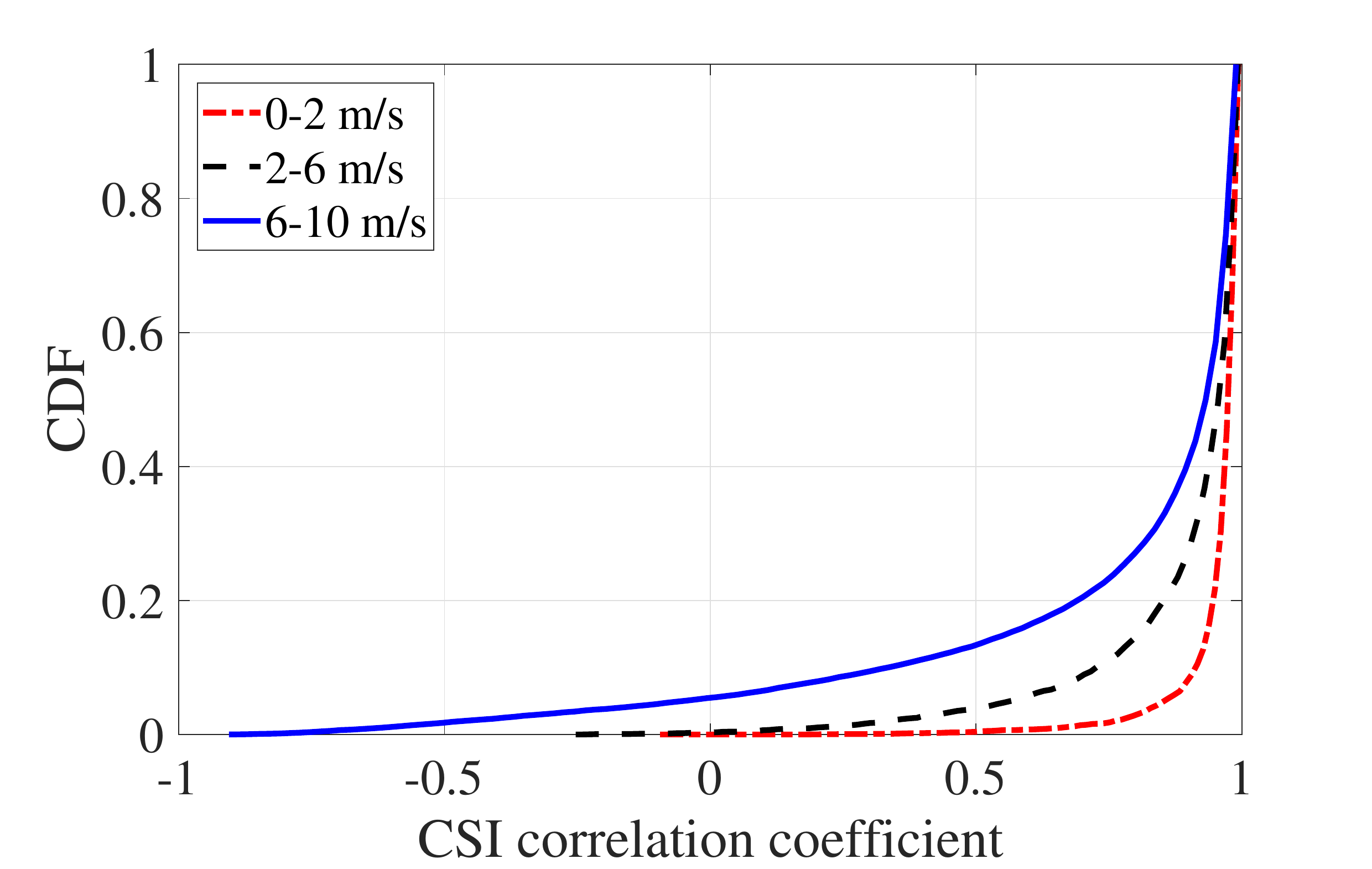}
		}
		\caption{Impact of different flight states.}
		\label{fig:state_impact}
	\end{minipage} 
	\begin{minipage}[b]{0.33\textwidth}
		\centering
		\includegraphics[width=0.95\textwidth]{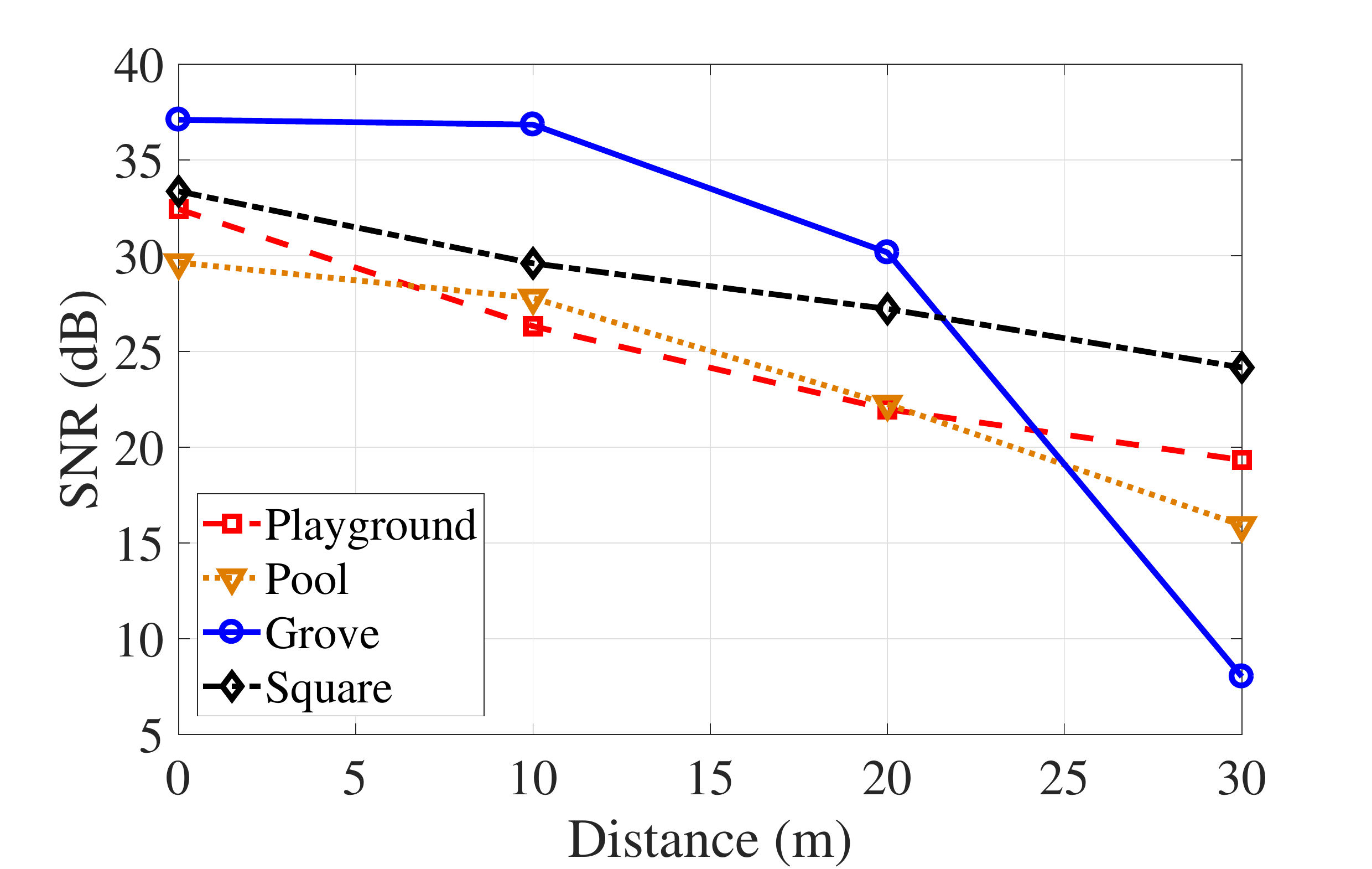}
		\vspace{0.5cm}
		\caption{Impact of different environments.}
		\label{fig:path_loss}
	\end{minipage}
	\vspace{-0.2cm}
\end{figure*}

To solve the above problems, our observation is that, different from the traditional wireless networks on the ground, the UAVs carry a wide variety of sensors to continuously derive the accurate flight state in real time.
That is, a UAV constantly estimates states including position, velocity, and orientation for autonomous navigation, which are crucial to maintain steady flight. 
We combine the sensor data that can reflect UAV states with the channel measurements in the rate adaptation algorithm to increase the communication throughput for air-to-ground links. 

In this paper, we propose \texttt{StateRate}, a rate adaptation system augmented with the state information extracted from the UAV's on-board sensors, as shown in Fig.~\ref{fig:schematic}. 
The goal of StateRate is to allow UAVs to predict optimal rates under dramatically varying flight states in unknown environments. 
StateRate estimates the current flight state from a variety of on-board sensors, and then combine the state and the channel information to infer the optimal MCS for the next frame. 
StateRate improves the throughput of air-to-ground links in dynamic channel conditions and flight states. 

There are two key challenges in StateRate to accurately predict the optimal rate. 
The first challenge is how to make full use of the information extracted from the sensor about the flight states. 
Traditionally, previous works~\cite{cars, sensor-hints} build a simple model based on the sensor data for indoor mobile and outdoor vehicular scenarios. 
Differently, the air-to-ground channel changes rapidly because UAVs are more flexible than vehicles.
Thus, the channel is more unpredictable and prohibitively complex to model. % freescatter 
To tackle this predicament, we treat the rate prediction as a multi-class classification problem and propose a prediction framework to choose the optimal rate. % RFID based, introduction
Specifically, we exploit convolutional layers to handle the frequency-selective fading and channel estimation error, and extract the temporal feature from the recurrent neural network (RNN). 

The second challenge stems from how to guarantee that the rate adaptation algorithm can adapt to a broad ranges of dynamic environments while we can only collect a limited amount of data in only a few environments.
To solve this problem, we propose an online learning algorithm to fine-tune our prediction framework when the performance of StateRate decreases in various environments. 
Specifically, we first exploit the difference between the path loss from multiple environments. Then we pre-train the network with a set of training data which is collected from a specific environments.
StateRate finally collects traces and is fine-tuned online to adapt to different environments.

The contributions are summarized below. % (T-Fi introduction) 
\begin{itemize}
	\item 
	We explore the impact of different UAV’s states and environments on channel conditions. 
	We develop a synchronization algorithm based on channel reciprocity to preprocess sensor data and channel measurements. 
	% We develop a model to capture the time-frequency features to support stable air-to-ground transmissions. 
	
	\item
	We design a rate adaptation algorithm based on on-board UAV sensors and propose a deep learning framework to dynamically predict the optimal rate. 
	In addition, we present an online learning algorithm to fine-tune the network parameters to adapt to various environments.
	
	\item 
	We implement our algorithm on a commercial UAV platform and compare our design with state-of-the-art rate adaptation algorithms under a wide range of conditions. 
	Experiment results show that StateRate improves the throughput up to 53\% compared with the best-known algorithm at a velocity of 2-6~$m/s$.
	
\end{itemize}

% (T-Fi introduction) 
The remainder of the paper is organized as follows. 
Section ~\ref{sec:motivation} analyzes the impact of different flight states and dynamic environments on the channel quality. Section~\ref{sec:design} introduces the system design of our rate adaptation algorithm. Performance evaluation of the system are elaborated respectively in~\ref{sec:evaluation}. Section~\ref{sec:relate} reviews related works and Section~\ref{sec:discussion} discuss the future work, followed by the conclusion in Section~\ref{sec:conclusion}.

\begin{figure*}
	\centering
	\subfigure[\scriptsize RSSI difference between adjacent packets]
	{
		\label{fig:AG_snr_difference}
		\includegraphics[width=0.45\textwidth]{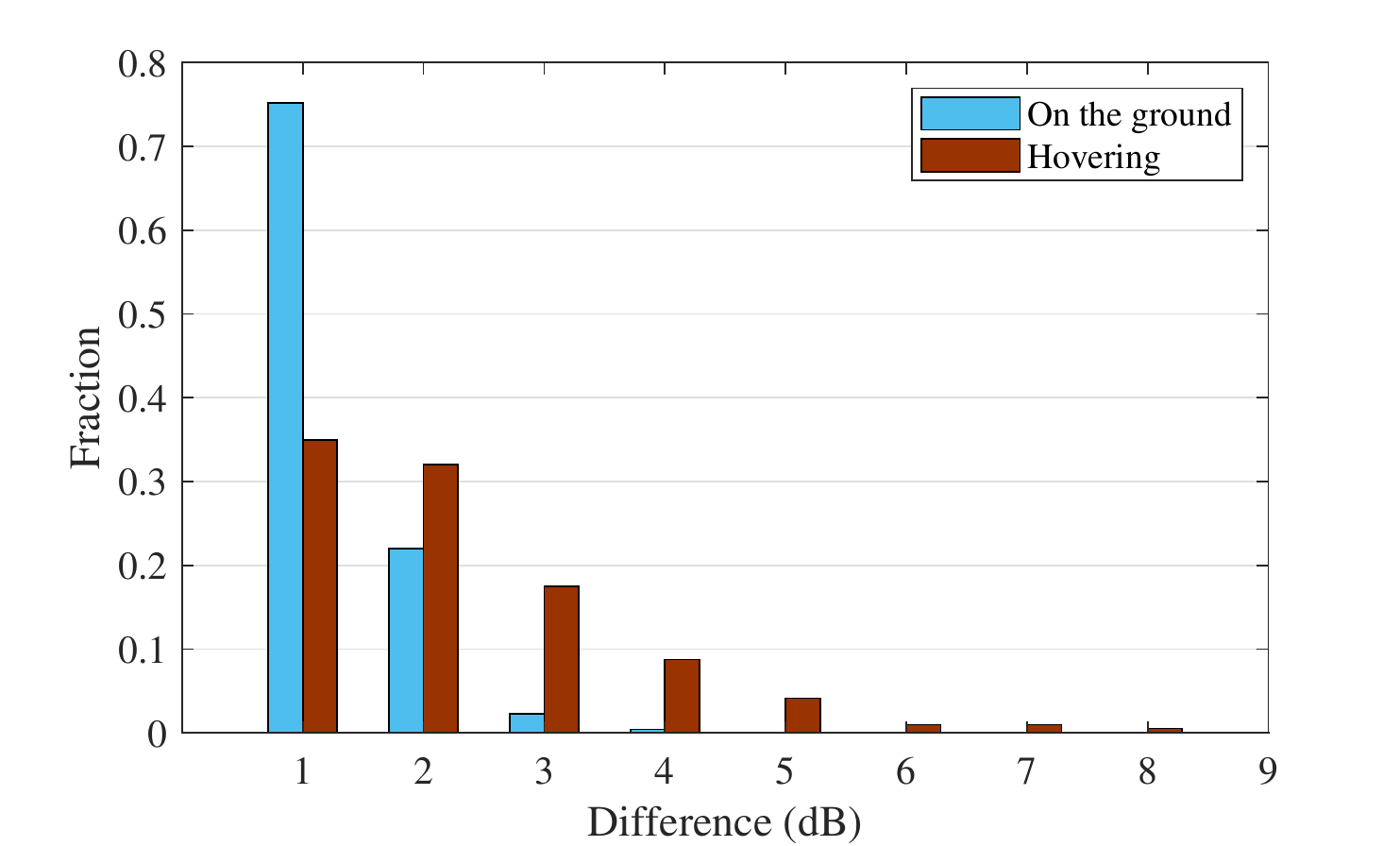}
	}
	\hspace{-0.2cm}
	\subfigure[\scriptsize CSI correlation of the two types of links.]
	{
		\label{fig:AG_csi_correlation}
		\includegraphics[width=0.45\textwidth]{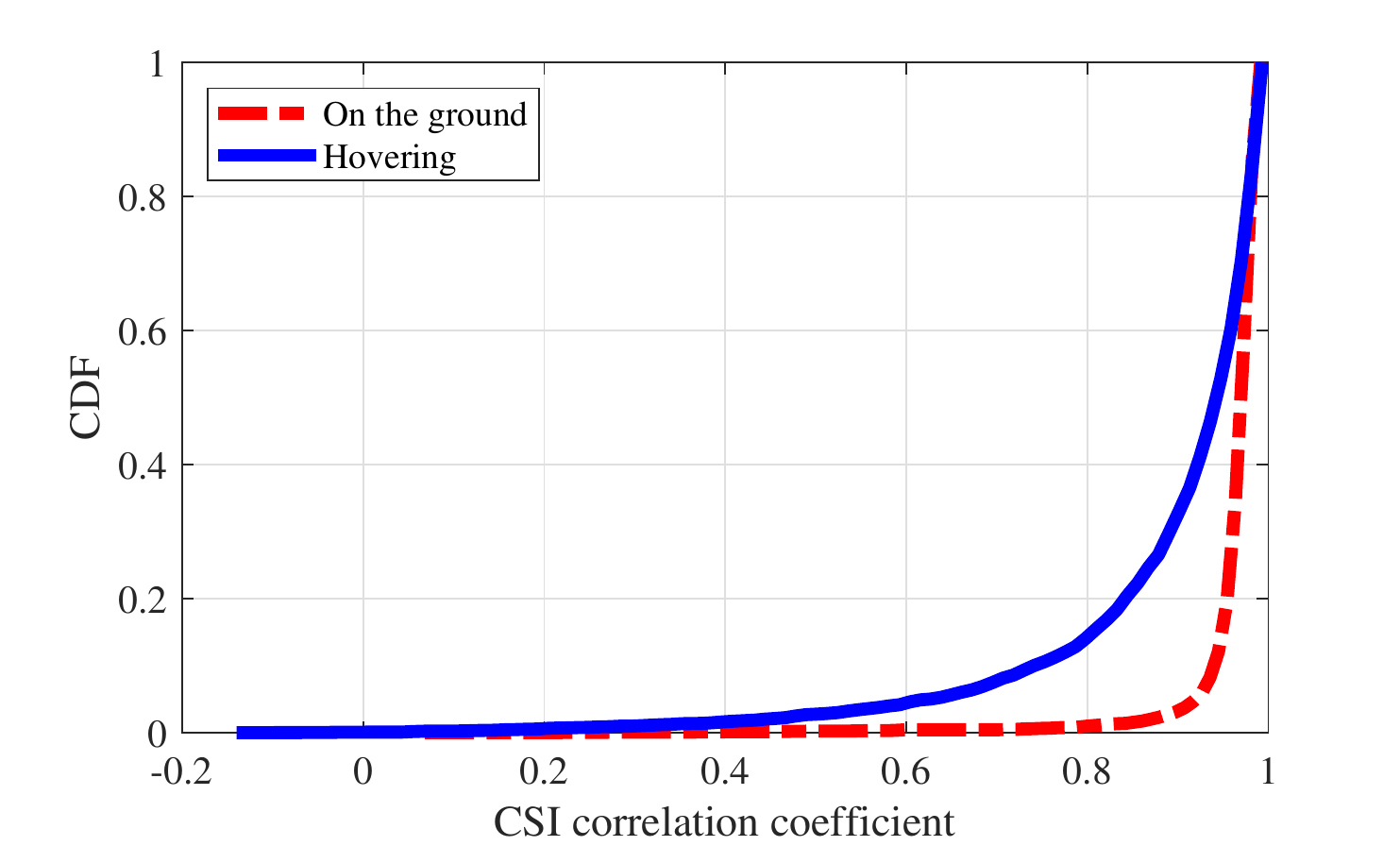}
	}
	\caption{Impact of UAV drift. The distance between the transmitter and the receiver is around $15~m$.}
	\label{fig:AG_impact}
\end{figure*} 

\section{Exploiting On-Board Sensor Opportunities}
\label{sec:motivation}

In this section, we first briefly introduce the background of state estimation in UAV and then elaborate on our observations of the impact of flight states on the air-to-ground channel. 

\subsection{UAV State Estimation Primer}

State estimation derived by the on-board sensors is an essential step in controlling UAVs for autonomous navigation. 
The goal of state estimation is to obtain a robust estimation of a UAV's flight state, including 3D position, 3D orientation, and their derivatives. 
For example, the position and velocity of a UAV are important parameters to tell whether it is hovering or moving. 
To change the state of motion, a UAV adjusts its orientation to generate a thrust.
In addition, in order to adjust the orientation, a UAV also needs to measure the angle and angle velocity in real time. 

Inertial measurement unit (IMU) and GPS are the sensors most commonly used to obtain the flight states in 3D space for a UAV. 
The angular velocity and acceleration can be directly obtained in IMU. 
However, due to the limited precision of IMU, the error is continuously accumulated. 
In the autonomous navigation systems for UAVs, the acceleration obtained from the IMU is generally fused with a GPS to ensure the accuracy of the velocity and acceleration through a long flight~\cite{GPS_IMU_Fusion}. 
Therefore, the flight states can be directly obtained from the fused data based on the on-board sensors (i.e., IMU and GPS).

\subsection{Observation}

The motion patterns for UAV are different from those on the ground. 
A rate adaptation algorithm needs to choose optimal rates that keep pace with flight states and environments. 
Traditional algorithms for mobile scenarios simply switch among multiple strategies based on whether the transceivers are mobile or static. 
Although these algorithms may work well in indoor mobile and outdoor vehicle environments, they may not be suitable for a more flexible air-to-ground communication system. 
The reason is the uncertainty caused by the rapidly changing flight states of the UAVs and the complicated communication environments.
Thus, it is necessary for a rate adaptation algorithm to design a framework that can make full use of the channel and sensor data.

% pensieve, learning abr algorithm
To motivate learning-based rate adaptation algorithms, we now provide some examples of where the flight states and environments affect the channel conditions in UAV scenarios. 
The experimental platform of the examples is composed of a DJI M100 UAV as the transmitter and a WARP v3 platform~\cite{warp} as the receiver. A laptop is connected to the WARP to collect channel traces, and the sensor data related to the flight states are extracted from the smartphone attached to the UAV. 
% Based on the results in the examples, we further summarize the direct motivation of our design due to the difference between aerial and ground communications.

% Pensieve, example; T-Fi experimental setup; mobirate, velocity-based loss-rate estimation; Mobility hints, Classifying Mobility using CSI
\textbf{Impact of flight states.}
We first conduct a motivational experiment to illustrate the impact of different flight states to the wireless channel. 
We control a UAV to randomly fly around the receiver and sweep the velocity from 0 $m/s$ to 10 $m/s$. Note that flying in a random trajectory is a more general setting to analyze the communication performance~\cite{random_trajectory}.
We test RSSI difference to indicate the time feature and CSI similarity to indicate the feature of frequency-selective fading, respectively.
During the experiment, the UAV sends a total of more than 30,000 packets.
Fig.~\ref{fig:snr_difference} shows the difference between the RSSI of two adjacent received frames from the UAV.
As expected, the RSSI difference between two adjacent frames increases with a higher velocity.
The fluctuation of the RSSI at high-velocity motion prevents traditional techniques from predicting optimal rates. 
% To further examine the impact of flight states on frequency-selective fading, we measure the CSI similarity under three different velocity scenarios.
Similarly, Fig.~\ref{fig:csi_correlation} shows the impact of flight states on frequency-selective fading.
In static scenarios, the CSI similarity stays close to one, while it drops under UAV mobility due to the rapid changes of the channel. 
This hampers existing algorithms to reach a high throughput. 
% In addition, if the UAV is moving, the multipath components and Doppler effect can cause time-varying channel.
% Therefore, the UAV communication system is required to properly assess such fluctuations and timely switch to a higher transmission rate.

% Pensieve example; mobirate, motivation; RAM, issues with packet statistics-based schemes. 
\textbf{Impact of environments.}
The second experiment considers the impact of different environments.
We choose four environments with different path-loss exponents and multipath components to collect the traces. 
Fig.~\ref{fig:path_loss} shows RSSI values in an experimental run for four different outdoor places.
The observation from the figure is that the path loss differs significantly across various places.
{Since the UAV needs to work in unknown environments, it would be inaccurate to only use only the same prediction strategy to choose the optimal rate for transmission in different environments.}

\textbf{Impact of drift.}
The final experiment illustrates the impact of the UAV drift. 
Compared with the flight state in Fig.~\ref{fig:state_impact}, the drift is more unpredictable and uncontrollable. 
In this experiment, we first continuously transmit data while the UAV is stationary on the ground, and then control the UAV to hover in the air.
The distance between the transmitter and the receiver is around $15~m$. 
Fig.~\ref{fig:AG_snr_difference} shows that the RSSI changes more drastically when the UAV is hovering compared when it is stationary.
When the UAV is on the ground, the RSSI difference can reach a maximum of $4~dB$, which is enough to choose an adjacent transmission rate as the optimal value.
In contrast, the RSSI difference can reach $8~dB$, indicating that the algorithm needs to find the optimal rate in a large search range.
Fig.~\ref{fig:AG_csi_correlation} shows a similar conclusion, calling for a new rate adaptation design.

\textbf{Summary.} 
It is not enough to predict a rate by showing the simplified impacts of the flight states and environments. 
Specifically, it requires a high-order function for prediction due to the fast UAV's velocity and dynamic transition of the flight states, while the ground channel is predicted by a linear prediction model. 
On the other hand, it is also hard to describe the number and shape of reflectors in an unknown environment. 
Traditional algorithms use fixed rules, so the performance under such conditions cannot be guaranteed.
In contrast, deep learning is very suitable for dealing with such complex conditions by modeling a high-order non-linear function without any prior knowledge. 
It can take advantages of the temporal characteristics in the channel sequence.

\section{Learning State-Aware Prediction}
\label{sec:design}

In this section, we illustrate our StateRate design. 
We start by explaining the system overview.
We then describe the preprocessing of input and the rate adaptation for different flight states using a prediction network.
Next, we explain how StateRate adapts to various environments.
We finally introduce rate optimization by describing the training process.

% softrate, esnr, packet deliver model; mobirate, overview
\subsection{Overview}
\label{subsec:overview}

\begin{table}[t]
	\centering
	\small
	\caption{Notations}
	\label{table:terminology}
	\vline
	\begin{tabular}{c|p{0.6\columnwidth}}
		\hline
		\textbf{Variable} & \textbf{Description} \\
		\hline
		$ C_n \triangleq (\text{csi}_n, \text{rssi}_n)$ & Channel state from the $n$th frame.\\
		\hline
		$ S_n \triangleq (d_n, v_n, a_n)$ & UAV flying state when the $n$th frame is received.\\
		\hline
		$\text{rssi}_n$ & Received RSSI from the $n$th frame.\\
		\hline
		$\text{csi}_n$ & Received CSI from the $n$th frame.\\
		\hline
		$d_n$ & Distance between the transmitter and the receiver when the $n$th frame is received.\\
		\hline
		$v_n$ & Velocity of the UAV when the $n$th frame is received.\\
		\hline
		$a_n$ & Acceleration of the UAV when the $n$th frame is received.\\
		\hline
		$\text{EMCS}_n$ & Evaluation of the $n$th frame's MCS when the $n$th frame is received.\\
		\hline
		$\text{PMCS}_{n+1}$ & Prediction of the $(n+1)$th frame's MCS when the $n$th frame is received.\\
		\hline
		$L_n$ & Label of the $n$th frame. It is a $8\times1$ vector representing the eight transmission rates.\\
		\hline
	\end{tabular}\vline
\end{table}

% We first describe the design overview of StateRate.
% Our goal is to develop a model that can accurately predict the packet delivery rate and improve throughput.
% To simplify the description, we summarize the terminology in our system in Table~\ref{table:terminology}.

Our goal is to develop a model that can accurately predict the packet delivery rate and improve throughput.
The communication protocol is based on IEEE 802.11ac standard~\cite{standard}, which is commonly used in UAVs.
The notational conventions used in this paper are summarized in Table~\ref{table:terminology}.

\begin{figure}[t]
	\centering
	\includegraphics[width=0.5\textwidth]{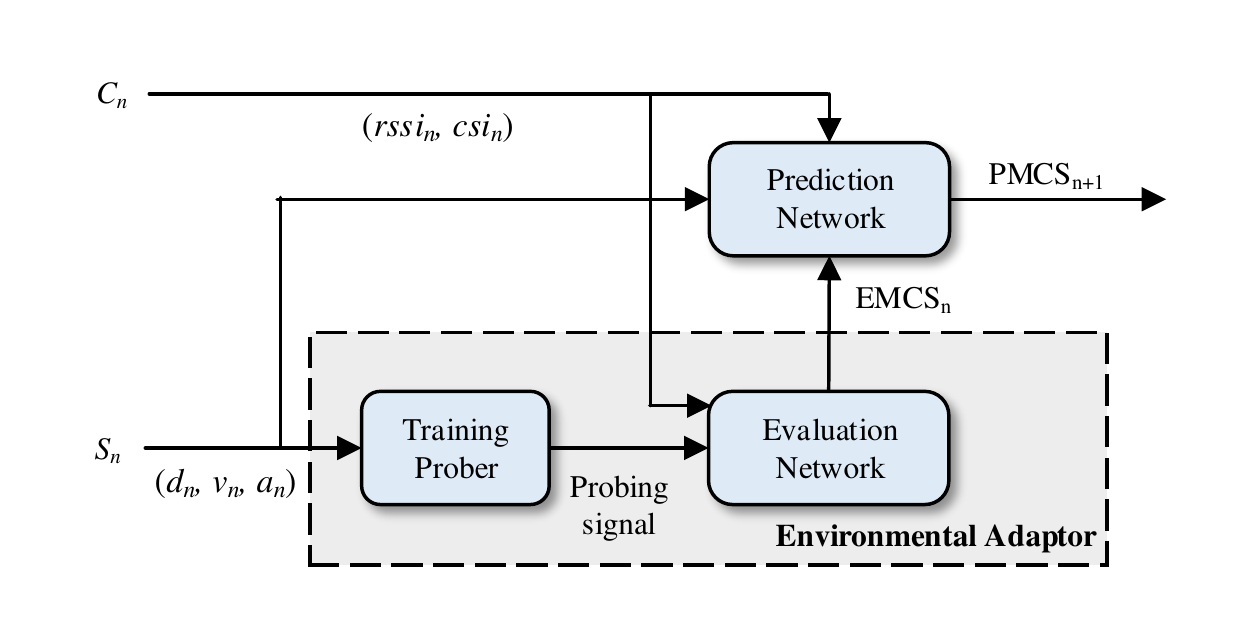}
	\caption{The framework of StateRate.}
	% \vspace{-0.2cm}
	\label{fig:overview}
\end{figure}

StateRate uses an SNR-based prediction method as its core. 
The primary distinction of this prediction process is that it is not a fixed but rather is flight states based. 
Second, it introduces an online learning framework that uses insufficient training data collected from unknown environments to enable environment independence. 
Finally, it has an environment-assisted prober that uses distance and prediction accuracy to trigger environmental adaptation. 
As shown in Fig.~\ref{fig:overview}, StateRate consists of three components: prediction network, evaluation network, and training prober.

\begin{itemize}

	\item \textbf{Prediction network.}
	Analogous to conventional SNR-based rate adaptation protocols, StateRate reads channel states including the $\text{csi}_n$ and $\text{rssi}_n$ from the $n$th frame and aims to predict the $\text{PMCS}_{n+1}$ of the $(n+1)$th frame. 
	The transmitter in UAV sends the $n$th data frame to the receiver on the ground.
	After the receiver detects the $n$th data frame, an ACK frame is sent back to the UAV immediately. 
	The UAV measures the channel information exported by the ACK frame. 
	The prediction network then combines the measured channel states from the physical (PHY) layer and the flight states from sensors to predict $\text{PMCS}_{n+1}$ for the $(n+1)$th frame. 

	\item \textbf{Evaluation network.}
	To adapt to different environments, the evaluation network acquires channel information from the ACK frame.
	This module outputs the evaluation MCS of the $n$th frame, i.e., $\text{EMCS}_n$.
	$\text{EMCS}_n$ is then fed to the prediction network to evaluate whether $\text{PMCS}_n$, which is predicted from the $(n-1)$th frame, is accurate or not.
	The evaluation network finally helps the timely fine-tuning of the parameters in the prediction network. 

	\item \textbf{Training prober.}
	The training prober monitors the environmental changes and determines when to fine-tune the prediction algorithm. 
	Once the training prober is triggered, the parameters in the prediction network would update online.

\end{itemize}

\begin{figure*}
	% \centering
	\begin{minipage}[b]{0.45\textwidth}
		\centering
		\includegraphics[width=0.95\textwidth]{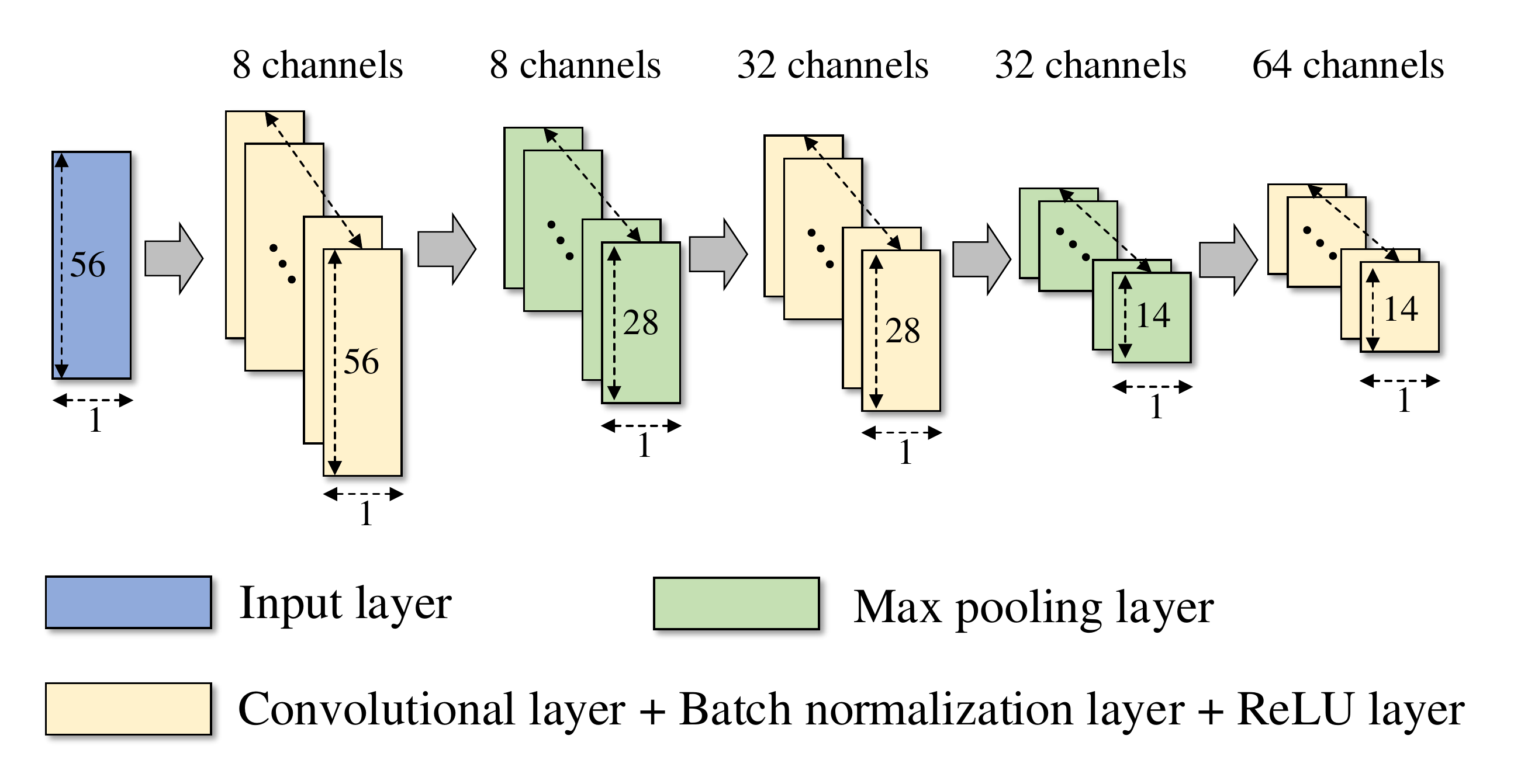}
		\caption{The CSI feature extractor in StateRate.}
		\label{fig:conv}
	\end{minipage}
	\hspace{0.9cm}
	\begin{minipage}[b]{0.45\textwidth}
		\centering
		\includegraphics[width=0.95\textwidth]{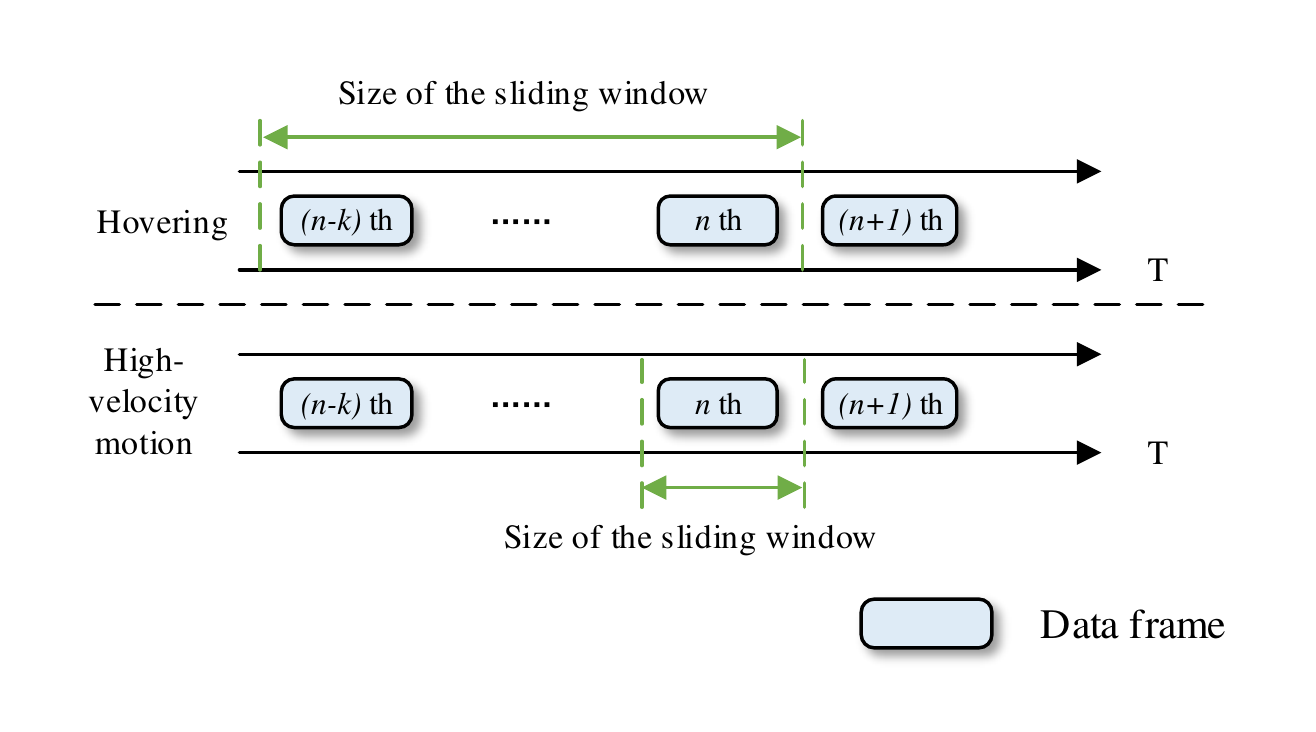}
		\caption{The size of the sliding window for different flight states.}
		\label{fig:sliding_windows}
	\end{minipage}
	% \vspace{-0.2cm}
\end{figure*}

\subsection{Preprocessing}

% Towards Environment Independent device, Section 2; Argos 2.4.1; Aerial Channel Prediction, Channel Prediction Algorithm
Before introducing the design of each module, we describe the preprocessing of input data. 
% Recall that we incorporate the sensor data into StateRate to provide hints about the rate prediction. 
% To this end, we consider how to integrate and synchronize two types of data.

Most of the SNR-based rate adaptation algorithms are deployed at the receiver because the channel state can be directly measured.
Unlike the conventional rate adaptation algorithms, StateRate combines $C_n$ and $S_n$ together to predict the $\text{PMCS}_{n+1}$. 
$S_n$ is obtained at the transmitter from the UAV, while $C_n$ is measured at the receiver on the ground. 
Thus, we need to integrate the input from both the transmitter and the receiver at the same time.

A straightforward idea is that the transmitter sends the sensor data to the receiver or the receiver feeds back the channel states to the transmitter at the lowest bit rate in real-time. 
% However, the cost of this scheme will be very expensive.
To solve this problem, we exploit channel reciprocity between the uplink and downlink to reduce the overhead.

Channel reciprocity is widely used in time division duplex (TDD) systems to estimate CSI implicitly~\cite{channel_reciprocity} and has been leveraged in rate adaptation algorithm~\cite{charm} to eliminate the need for RTS/CTS.
To ensure channel reciprocity works appropriately, we suppose that the UAV in StateRate moves at a maximum velocity of $v_m = 20~m/s$.
The maximum Doppler shift is given by
\begin{align}
	f_m = \frac{v_m}{c} \times {f_c}, \notag
\end{align}
with $f_c$ being the center frequency. The channel coherence time $T_m$ can be expressed as
\begin{align}
	T_m = \frac{0.423}{f_m} \simeq 2.644~ms. \notag
\end{align}
In IEEE 802.11 standard, the receiver sends an ACK frame after the short interframe space (SIFS), which is $10~\mu s$. 
The delay of the ACK frame is much smaller than the channel coherence time although the UAV moves at the fastest velocity, so channel reciprocity is appropriate for the UAV scenario.
After detecting the data frame, the receiver feeds an ACK frame back to the transmitter.
The transmitter on the UAV then uses the channel state of the ACK frame as the accurate estimation of the uplink channel. 

Another issue is the synchronization between the two types of data.
Although both the channel measurement and sensor data can be obtained at the transmitter, the sample rate of the sensor is still lower than the sample rate of the channel.
Therefore, one sensor measurement needs to correspond to multiple channel measurements.
To deal with this problem, we record the timestamp of each sensor data and synchronizes the channel measurement to the sensor data with the smallest difference in timestamps.

% \begin{figure}[t]
% 	\centering
% 	\includegraphics[width=0.47\textwidth]{figures/Conv}
% 	\caption{The CSI feature extractor in StateRate.}
% 	\vspace{-0.2cm}
% 	\label{fig:conv}
% \end{figure}

\subsection{Prediction Network}
\label{subsec:PN}

We propose a prediction network $P(\cdot)$ to pick the optimal rate and enable our algorithm to adapt to different flight states in the same environment. 
This network combines $C_n$ and $S_n$ together to predict $\text{PMCS}_{n+1}$.
The output of the network is a $8 \times 1$ vector representing the weight of eight MCSs, which can be expressed as
\begin{align}
	W^P_{n+1} = \text{Softmax}(P(C_n, S_n)),
\end{align}
where $W^P_{n+1}$ denotes the probability of the transmission rate.

% RSSI includes the power level of the received packets, which is the primary factor about the transmission rate. CSI contains the information about the frequency-selective fading. The more faded channels are more likely to increase bit errors. % ESNR, Impact of Frequency-Selective Fading
% Traditionally, the raw data of the CSI, expressed as $rcsi_n$, cannot be directly entered to the network because CSI are represented as several bits to give the value of real and imaginary parts at the receiver. 
% The CSI obtained in our experiments thus lacks of the energy information.
% In order to achieve the energy in $csi_n$, we normalize $rcsi_n$ with $rssi_n$ 
% \begin{align}
% 	csi_{nk} = \frac{rcsi_{nk}}{\sum_{k=0}^{63}{rcsi_{n}}} * rssi_n.
% \end{align}
% where $k$ indicates the index of the subcarriers. 

\begin{figure*}
	% \centering
	\begin{minipage}[b]{0.45\textwidth}
		\centering
		\includegraphics[width=0.8\textwidth]{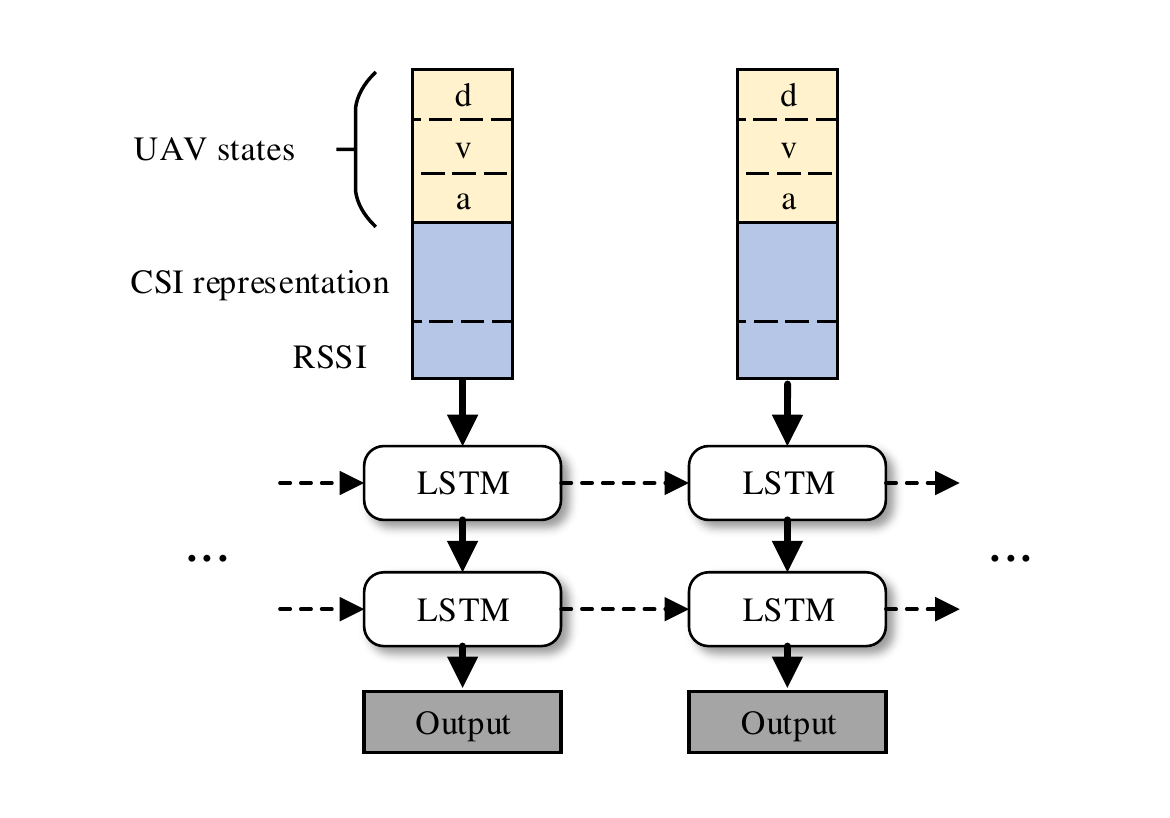}
		\caption{The prediction modeling.}
		\label{fig:lstm}
	\end{minipage}
	\hspace{0.9cm}
	\begin{minipage}[b]{0.45\textwidth}
		\centering
		\hspace{-0.2cm}
		\includegraphics[width=0.97\textwidth]{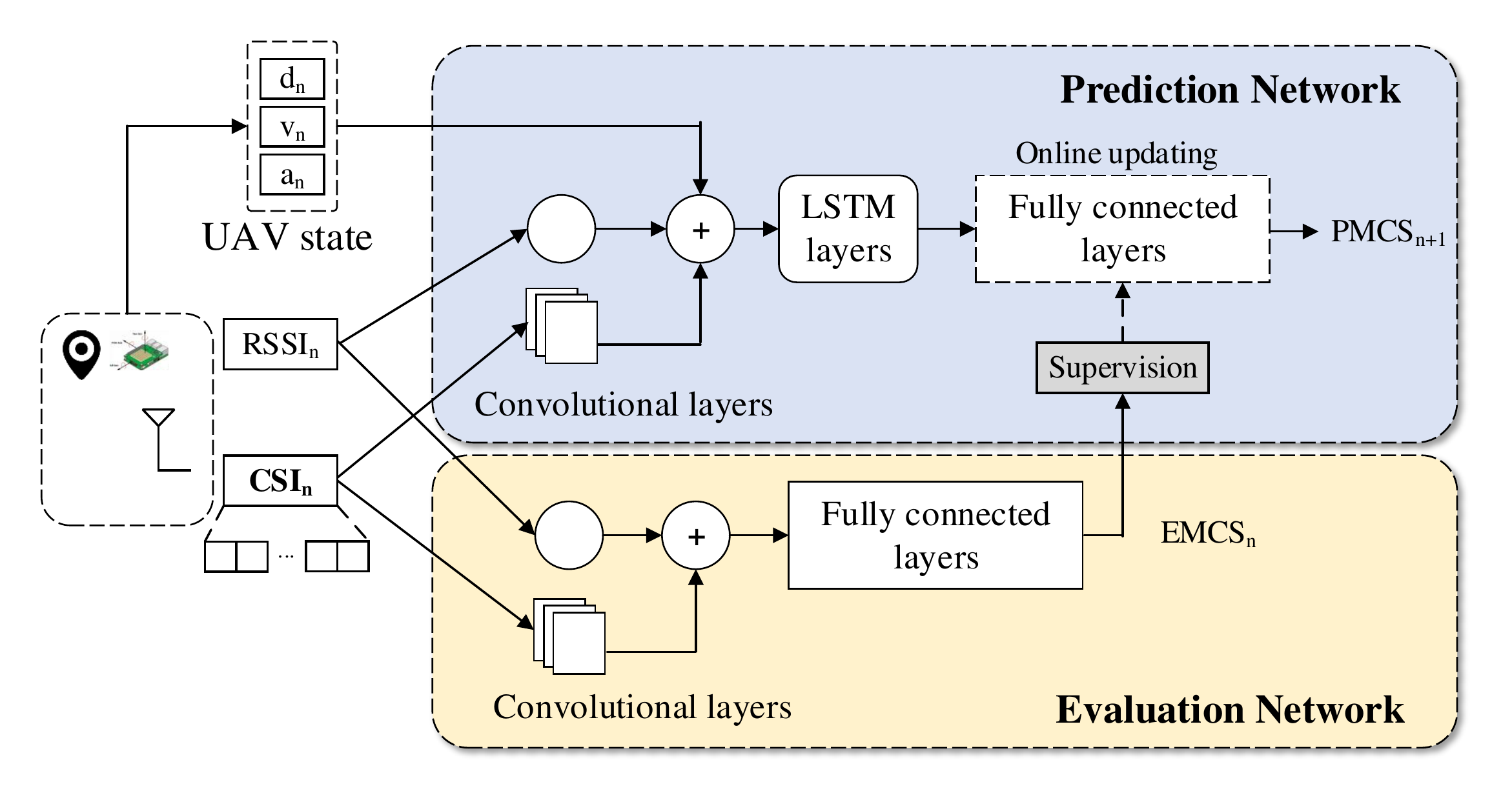}
		\caption{ The online training in StateRate.}
		\label{fig:network}
	\end{minipage}
	\vspace{-0.1cm}
\end{figure*}

% \begin{figure}[t]
% 	\centering
% 	\includegraphics[width=0.46\textwidth]{figures/Online_Training}
% 	\caption{ The online training in StateRate.}
% 	\label{fig:network}
% \end{figure}

\textbf{CSI feature extractor.}
The second problem to consider is what kinds of channel parameters are used in the prediction component and how to make full use of the information.
RSSI and CSI have already been implemented in the off-the-shelf network card~\cite{tool_release}. 
Thus, we choose RSSI and CSI as our parameters without modifying the existing hardware. 
ESNR\cite{esnr} is the most common metric based on the RSSI and CSI to represent the channel, while it is still insufficient to track the channel due to the channel estimation error. 
% In addition, although mapping SNR to BER is determined, it is hard to obtain the exactly BER when the SNR is known theoretically. 
% We need a more accurate metric to describe the relationship between SNR and BER.
% Another approach is to set a smoothing window to measure CSI more accurately. 
% However, the size of the smoothing window is difficult to determine due to the dynamic flight states. 

To address this problem, we use a learning-based feature extractor to track accurate channel information.
We employ convolutional neural networks (CNNs) to extract the frequency-selective fading features. % environment independent HAR; feature extractor
As shown in Fig.~\ref{fig:conv}, it consists of three convolutional layers and takes $\text{csi}_n$ as input and returns the feature representation matrix. 
After each convolutional layer, some components are introduced to optimize the training process.
In particular, we use batch normalization layers directly after each layer to standardize the distribution of the output~\cite{batch}.
The activation functions ReLU are then concatenated to increase the sparsity of the network.
We finally use the pooling layer to reduce the parameters of the first two convolution layers.
These components reduce the convergence time and prevent over-fitting.

\textbf{Prediction modeling.}
Rate adaptation is a predictive function, so we propose a prediction module for temporal modeling and prediction.
Traditionally, the prediction model is based on past channel measurements with a sliding window. 
As shown in Fig.~\ref{fig:sliding_windows}, a large sliding window is necessary when the channel changes slowly, and a small one is required when the channel changes rapidly. 
We consider that the prediction module needs to quickly adapt to the flight state, so it is difficult to determine the size of the sliding window in real-time. 

% spatiotemporal, temporal modeling and prediction
To solve the problem of the dynamic sliding window, we combine the last channel state and flight state with the internal memory of the Long Short Term Memory (LSTM) to process sequences of inputs.
LSTM is a common network of deep learning for modeling temporal sequence, such as modeling of precipitation nowcasting. 
Specifically, we employ an LSTM with two hidden layers for temporal modeling, as shown in Fig.~\ref{fig:lstm}. 
It takes the representations from the CSI feature extractor and the UAVs states as input. 
To predict the future optimal rate, the two types of inputs will be concatenated as a vector. 
The cell then combines the previous cell states and the current inputs to update the parameters in the hidden layers. 

\textbf{Classification layers.}
The classification layers are two fully connected layers, acting as the end block of our prediction model. 
It takes the output of the LSTM layers as input for classification and outputs a $8 \times 1$ vector representing the eight transmission rates. 

\subsection{Environmental Adaptor}
\label{subsec:env_adapt}

In this section, we illustrate how StateRate adapts to unknown environments. 

% RF-Pose, Cross-Modal Supervision
The environmental adaptor trains the prediction network online to ensure that StateRate can work in unknown environments.
The challenge of designing an environmental adaptor is that the collected traces lack labels in new environments. 
Ideally, the label can be directly mapped from $C_n$. 
The label is obtained once the $n$th frame is received. 
For example, at $n$th time slot, we obtain the current channel information, and the optimal transmission rate is determined at the same time by leveraging a fixed mapping. 
The prediction network needs to predict the optimal transmission rate at the $(n+1)$th time slot. 
When the next packet transmission is completed at the $(n+1)$th time slot, StateRate knows the label of $(n+1)$th transmission rate, and then it needs to predict the optimal rate at the $(n+2)$th time slot. 
However, since the parameters in the channel model is unknown, it is difficult to derive the closed expression between $C_n$ and $L_n$. 
To obtain the label, we change the MCS and repeatedly send data for each training traces. 
% We consider the MCS with the highest rate of $\text{PRR} > 90\%$ as the label.

The above scheme cannot be directly deployed into the environmental adaptor due to its significant overhead. It is almost impossible to annotate optimal MCS by collecting massive training data in a short flying time. To solve the this issue, we aim to design an evaluation network, whose goal is to give an accurate evaluation metric when StateRate runs in unknown environments.

\textbf{Evaluation network.} 
The evaluation network is the core part of the environmental adaptor. Similar to the prediction network, the evaluation network consists of a CSI feature extractor and classification layers. In contrast to the prediction network, the evaluation network only employs $C_n$ from the $n$th received frame as the input and outputs the probability of the evaluation MCS as the ``virtual label''. The expression can be described as
\begin{align}
	W^E_n = \text{SoftMax}(E(C_n)).
\end{align}
The evaluation network does not contain any information about the flight states as it does not need to predict the future rate. 
Thus the output is only related to the current channel information. 

\textbf{Training Prober.}
We design a training prober to determine when to trigger online training.
A straightforward idea is to fine-tune the network at any time, but this will lead to over-fitting if the network is continuously trained with the data from the same environment.
Besides, continuous training will also increase energy consumption and occupy computing resources.
Therefore, it is necessary to detect changes in the surrounding environment and to ensure that the online training only works when the environment changes.
We have two metrics to trigger online training: 1) the real-time distance of the UAV relative to the takeoff point and 2) the prediction accuracy.
Online training is triggered after the follow two steps:
\begin{enumerate}
    \item[1)] StateRate saves the channel data and sensor data collected in a new environment.
    \item[2)] The prober detects the environment change and triggers online training.
\end{enumerate}
It is worth noting that online training only updates the parameters in the network, while the channel does not update during the online training period.

% We further design a sliding window to estimate the prediction accuracy. 
% If the sliding window is large, it may contain data from different environments; if the sliding window is small, the data fluctuates greatly.
% The size of the sliding window is set to 1,000 packets in our experiment to strike a good balance.

% The channel traces and flight states which are collected in a new environment are used for fine-tuning the prediction network.
% To ensure the speed of adaptation to the environment, training data cannot be collected for too long. In order to reduce the speed of online training, the training dataset should not be too small.
% Therefore, in order to ensure that the environment is adapted in a short time, we choose 10,000 traces collected in the $20~s$ for online training.

\begin{figure*}
	\begin{minipage}[b]{0.45\textwidth}
		\centering
		\subfigure
		{
			\includegraphics[width=0.82\textwidth]{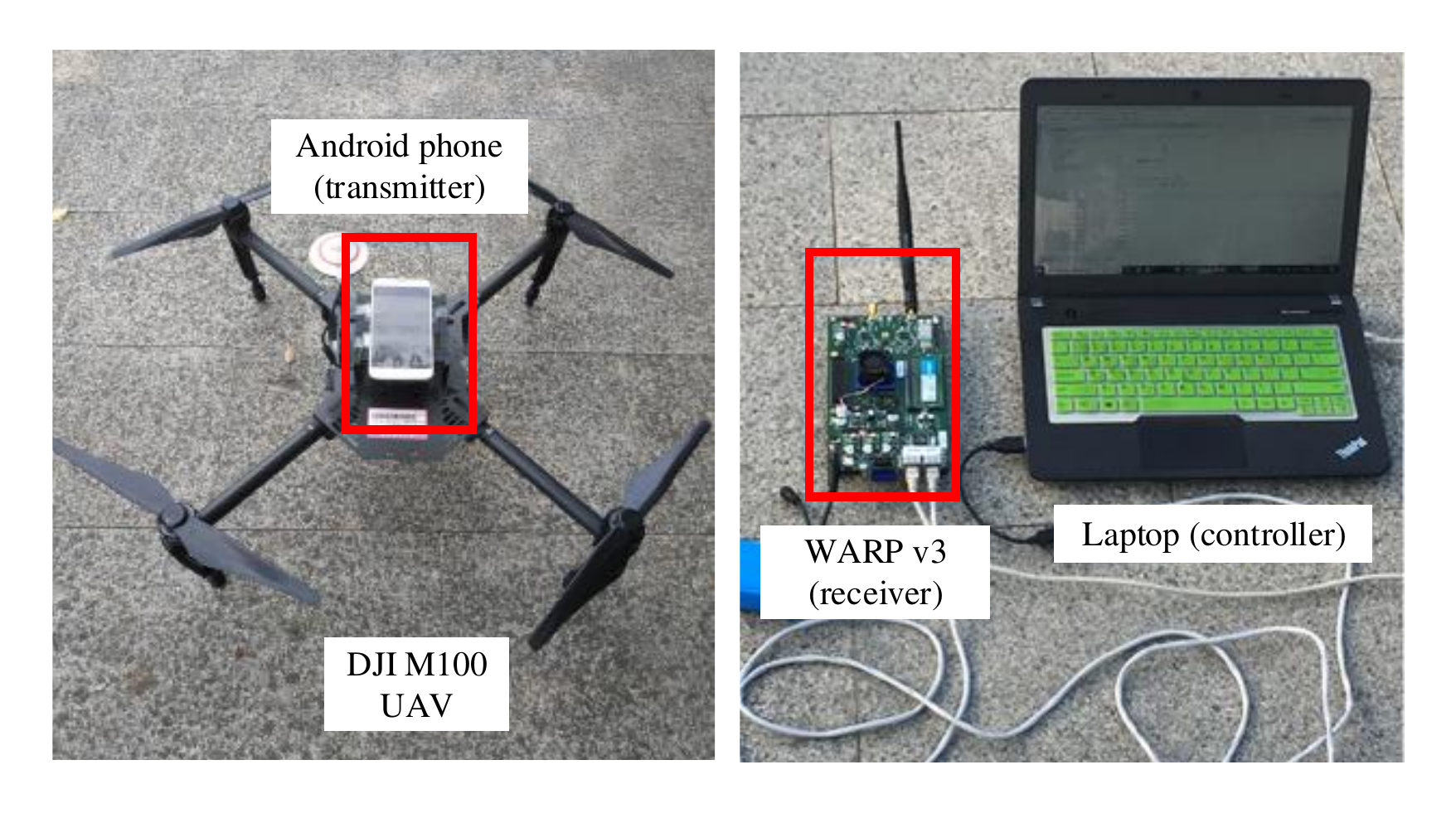}
		}
		\caption{Implementation.}
		\label{fig:Implementation}
	\end{minipage}
	\hspace{0.9cm}
	\begin{minipage}[b]{0.45\textwidth}
		\centering
		\subfigure
		{
			\includegraphics[width=0.9\textwidth]{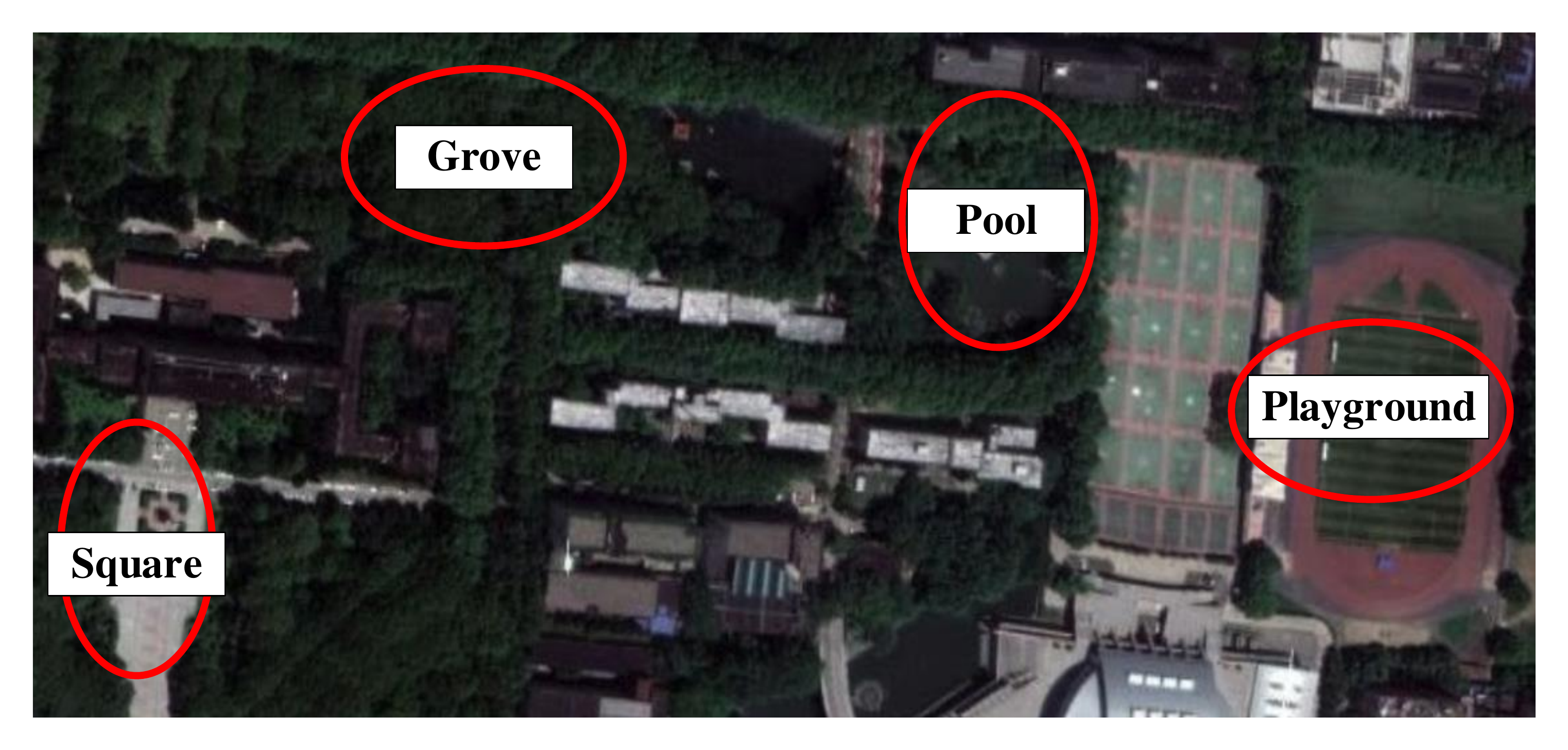}
		}
		\caption{Experiment sites.}
		\vspace{0.1cm}
		\label{fig:sites}
	\end{minipage}
	% \vspace{0.2cm}
\end{figure*}

\subsection{Rate Optimization}

Generally, the transmission rate is optimized by training the prediction network.
At a high level, we train an offline model using the collected training data from finite environments and then fine-tune the pre-trained model after the UAV flies into a new environment.

% Environment independent; Activity Recognizer
\textbf{Offline rate optimization.} 
We use the cross-entropy function to calculate the loss between the $n$th predictions and the labels as
\begin{equation}
	\text{Loss}_p = \sum_{i=0}^{7}{l_{n, i} \log{w^P_{n, i}}},
\end{equation}
where $l_{n, i}$ and $w^P_{n, i}$ are the element of $L_n$ and $W^P_n$, respectively.
The optimization process is based on backpropagation and the parameters in the network are iteratively updated.
Besides, the evaluation network can also be trained offline. The loss function can be expressed as
\begin{equation}
	\text{Loss}_E = \sum_{i=0}^{7}{l_{n, i} \log{w^E_{n, i}}}.
\end{equation}
These two well-trained networks will serve as a pre-trained model for online learning.

\textbf{Online rate optimization.} 
We use the cross-modal supervision~\cite{crossmodal} in which the output of the prediction network is compared with the virtual labels to train our model, as shown in Fig.~\ref{fig:network}.
The training objective of the online training process is to minimize the difference between the two networks. The loss function is
\begin{align}
	\hat{\text{Loss}}_P = \sum_{i=0}^{7}{w^E_{n, i} \log{w^P_{n, i}}}.
\end{align}

In order to speed up the real-time fine-tuning of the network, the online training is slightly different from the offline training.
Specifically, if the environment changes, the prediction network is trained on the basis of the pre-trained model. 
We only fine-tune the parameters in the fully connected layers of the prediction network.

\section{Evaluation}
\label{sec:evaluation}

In this section, we conduct a set of detailed experiments to show the advantages of our design under various flight states and environments.

\subsection{Experiment Setup}
\label{sec:implement}

\textbf{Implementation.}
% To evaluate StateRate under realistic channel conditions, we build an experimental platform to collect sensor data and channel states. 
The platform, as shown in Fig.~\ref{fig:Implementation}, is composed of a DJI M100 UAV, an Android mobile phone, a WARP v3 testbed, and a laptop.
We collect two types of data, one is the sensor data of the UAV, and the other is the real-time channel traces.
Since we need to synchronize the channel traces and the sensors, we fix the Android phone to the UAV and obtain the sensor data from it.
The Android phone sends packets to a wireless AP on the ground.
The AP is a WARP v3 board, which can run the IEEE 802.11 protocol independently.
WARP is connected to a laptop through an Ethernet connection. 
The channel information and sensor data are synchronized with the received timestamp in WARP.
In addition, the network is trained in a desktop with an 8-core 64-bit Intel i7-6850k processor, a GTX 1080 Ti GPU, and a 32GB RAM. In the above setup, the training time of 112,000 traces is 15 minutes and 11 seconds when we run 100 epochs.

The experiment sites are shown in Fig.~\ref{fig:sites}.
In order to evaluate the performance in different cases, random trajectories of the UAV are used for each test.
We collected a total of 570,000 channel traces and corresponding sensor data from four environments (a square, a playground, a pool, and a grove) to evaluate our algorithm.
It is worth noting that we conduct experiments in real-world environments, where there exists interference from other links. To make it compatible with IEEE 802.11 protocols, we conform to the legacy CSMA/CA mechanism to address the interference problem.

% ESNR trace-driven simulator
\textbf{Offline data processing.}
To compare the algorithm with state-of-the-art rate adaptation algorithms under the same UAV conditions, we turn to offline data processing to implement our algorithm.
In particular, the collected CSI and RSSI serve as the ground truth for the offline processing.
We implement the packet transmission pipeline, including modulation and the channel coding modules.
The transmission rates used in our experiments are BPSK, QPSK, 16QAM with code rates of $1/2$ and $3/4$, and 64QAM with code rates of $2/3$ and $3/4$, which are widely used in IEEE 802.11 standards.
Ideally, the MCS is determined to maximize the throughput, which is related to both the transmission rate and packet loss. Moreover, packet loss is as small as possible to reduce the retransmission overhead. Therefore, we choose the optimal MCS when the loss rate is less than 10\%. 

\begin{figure*}
	\centering
	\subfigure[\scriptsize Accuracy]
	{
		\label{fig:overall_accuracy}
		\includegraphics[width=0.45\textwidth]{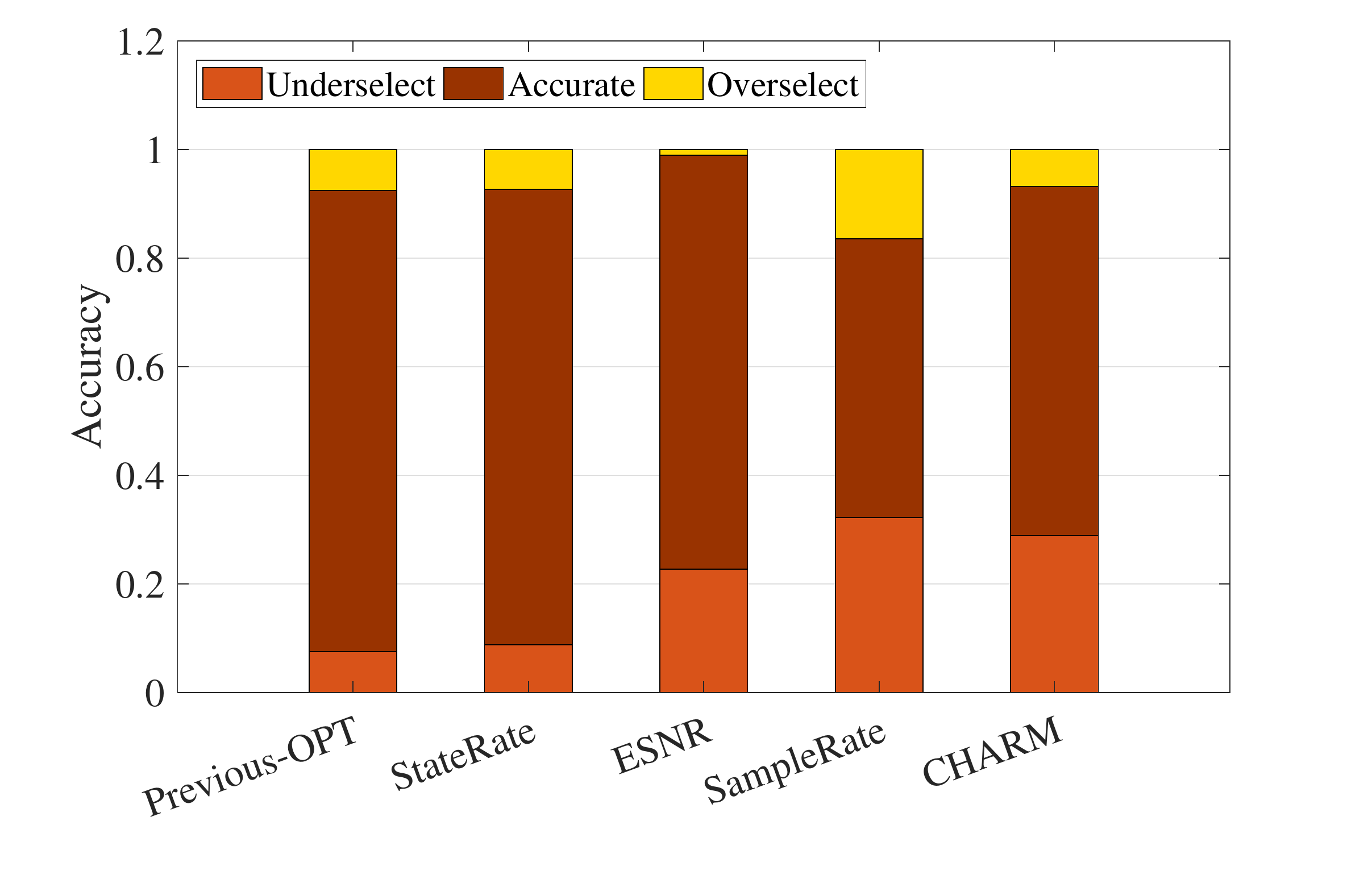}
	}
	\hspace{-0.2cm}
	\subfigure[\scriptsize Throughput]
	{
		\label{fig:overall_throughput}
		\includegraphics[width=0.45\textwidth]{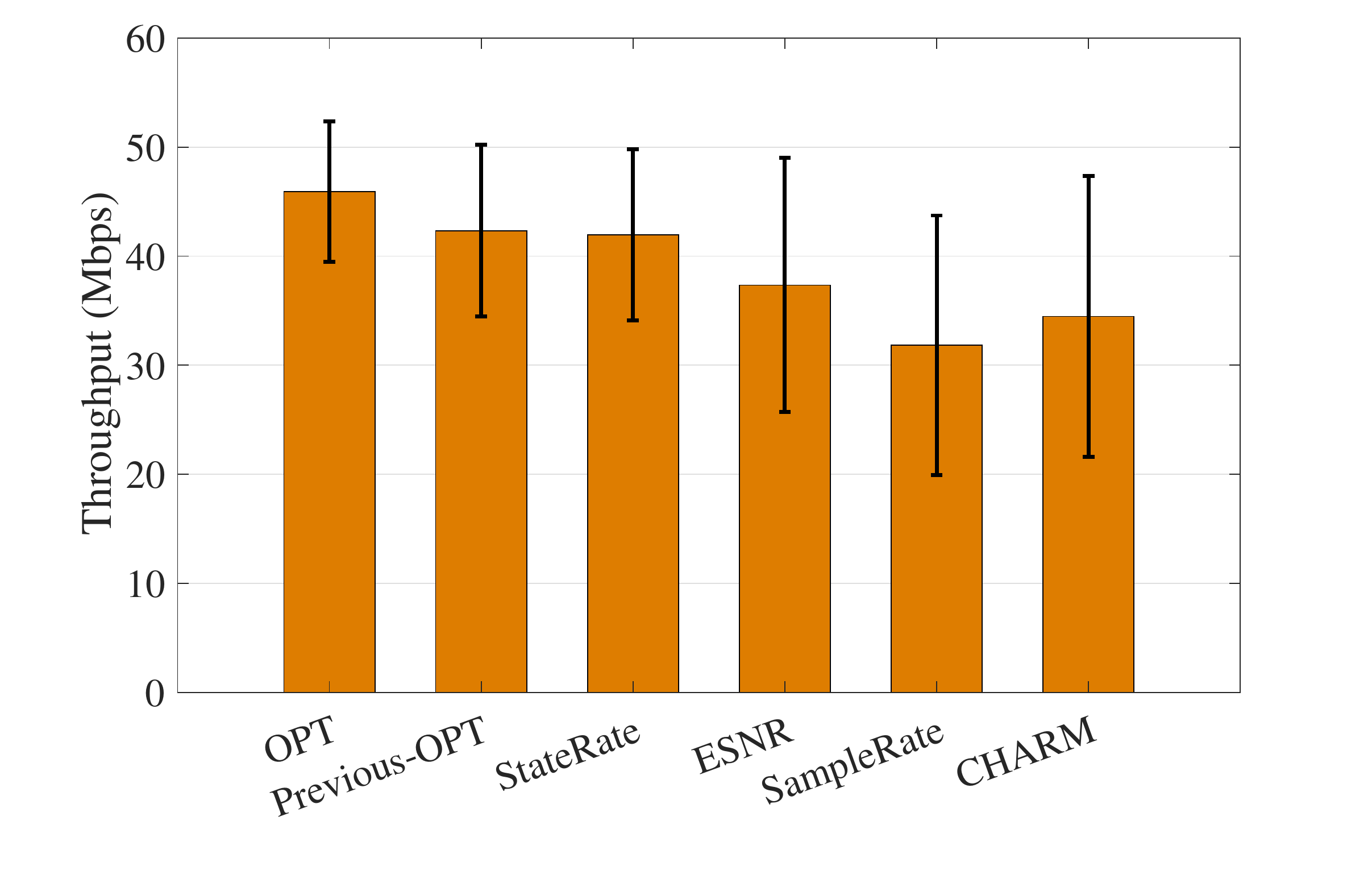}
	}
	\caption{Overall performance.}
	\label{fig:overall_performance}
	% \vspace{0.4cm}
	\vspace{-0.1cm}
	\centering
	\subfigure[\scriptsize Accuracy vs velocity]
	{
		\label{fig:accuracy_evaluation}
		\includegraphics[width=0.45\textwidth]{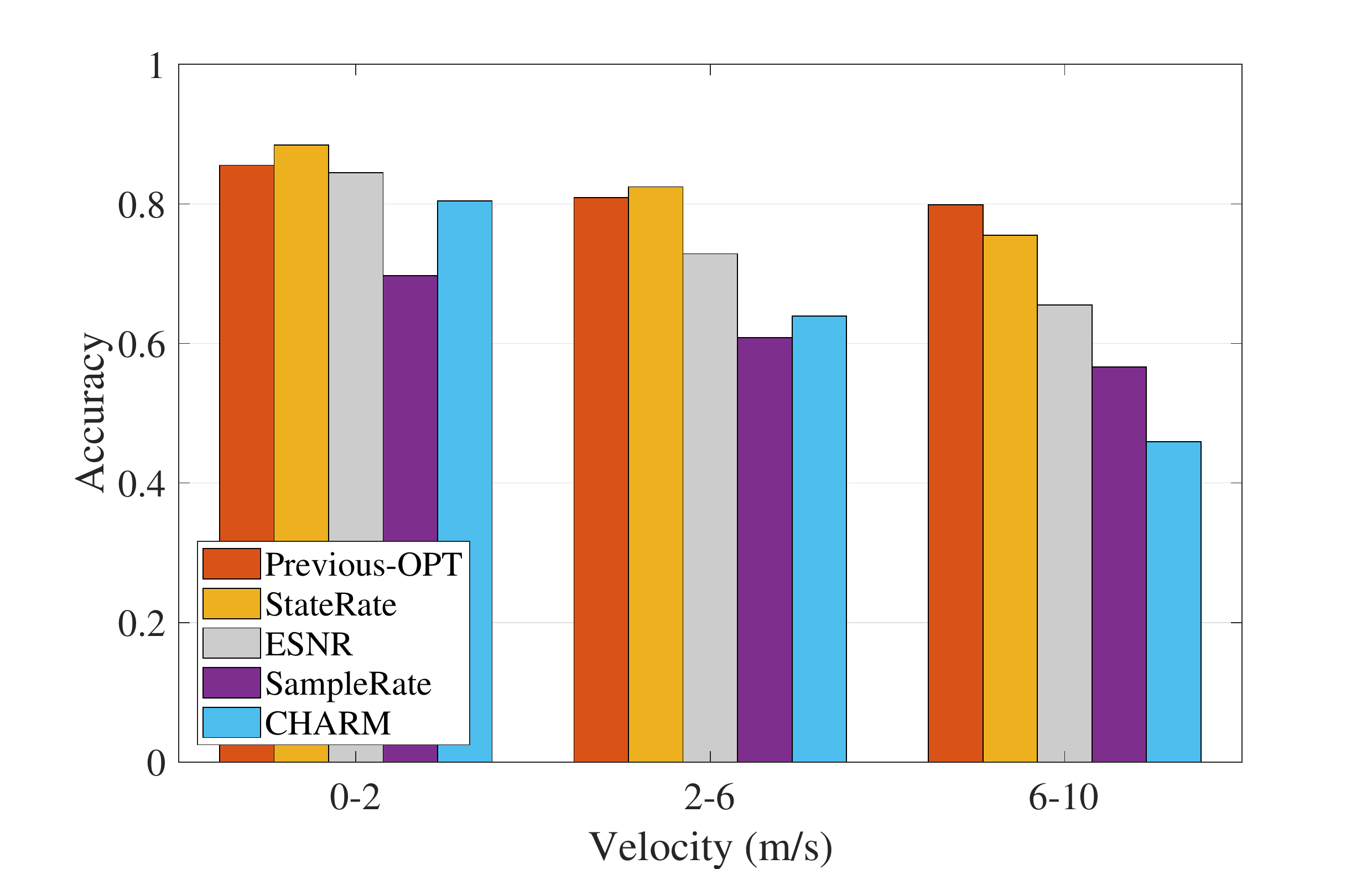}
	}
	\hspace{-0.2cm}
	\subfigure[\scriptsize Throughput vs velocity]
	{
		\label{fig:throughput_evaluation}
		\includegraphics[width=0.45\textwidth]{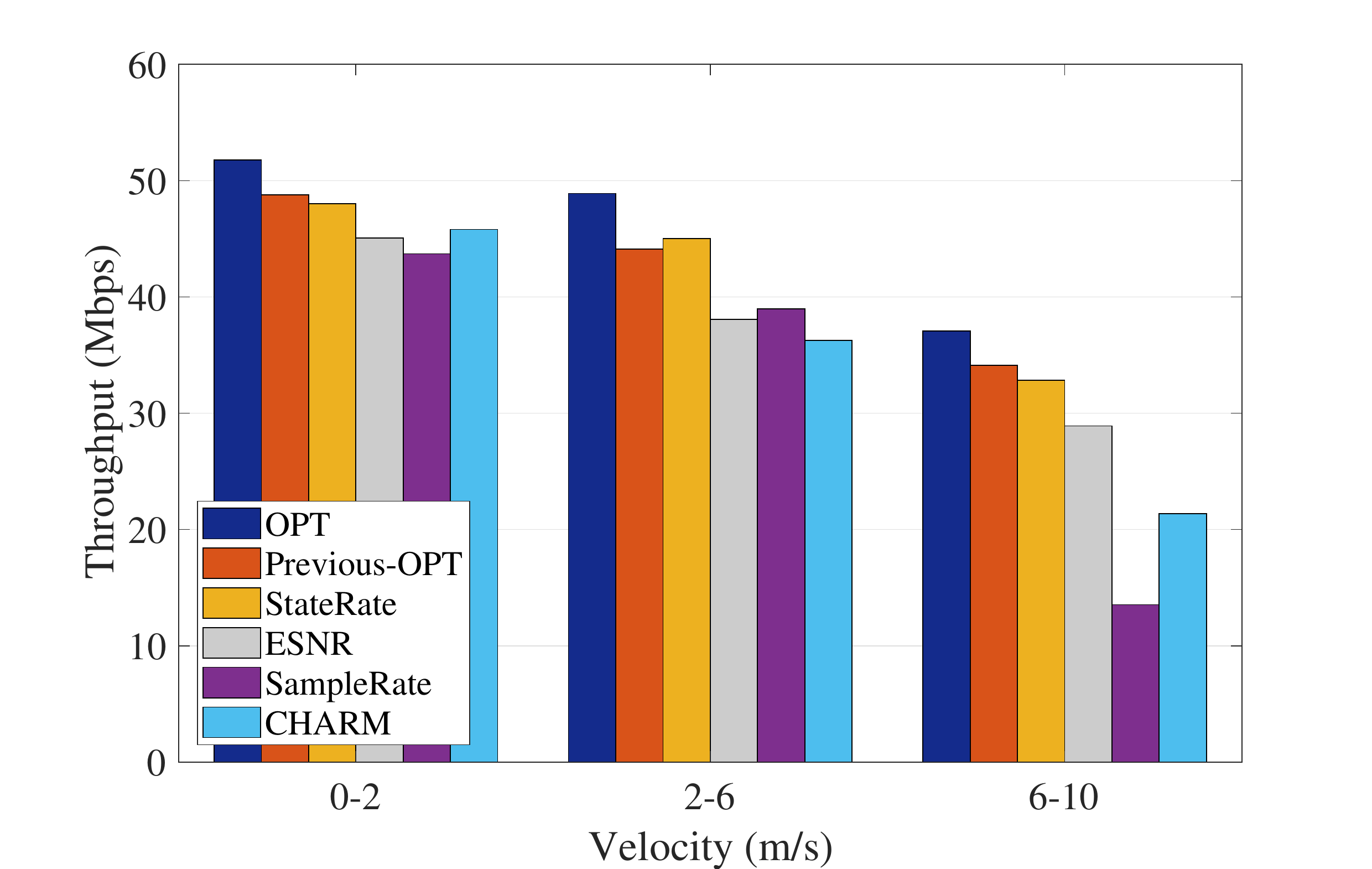}
	}
	\caption{Performance under different velocities.}
	
	\label{fig:states_evaluation}
\end{figure*}

% softrate, evaluation of softrate
\textbf{Baseline algorithms.}
% We evaluate StateRate using trace-driven emulation as described in Section~\ref{sec:implement}.
We compare StateRate against other rate adaptation algorithms using two matrics: prediction accuracy and throughput. 
% \hl{While previous works mostly only use throughput as a measure of performance, we believe that the prediction accuracy obtained from the neural network significantly affects the throughput in most scenarios.}
The accuracy ($R$) is obtained by comparing the prediction result ($\text{PMCS}_n$) with the label ($L_n$), i.e., $R = N_p / N$, where $N_p$ and $N$ are the number of correct prediction results and the number of all traces, respectively.
We experiment with StateRate and three representative algorithms as well as two omniscient algorithms.
These representative algorithms can be readily applied to commercial devices, including two categories, i.e., loss-based algorithms (SampleRate) and SNR-based algorithms (CHARM, ESNR).

\begin{itemize}

	\item
	\textit{SampleRate}~\cite{samplerate} is a loss-based algorithm that uses packet loss and transmission time to choose the bit rate. 
	It sends a packet to probe a random rate every ten packets and switches to it only if the transmission time is greater than the time of the current rate. 
	On the other hand, it decreases to a lower bit rate when the transmission time of the lower rate is less than the time of the current rate. % esnr, rate selection algorithm

	\item
	\textit{CHARM}~\cite{charm} is an SNR-based rate adaptation algorithm that uses signal strength to choose the transmission rate. 
	It proposes a weighted sliding window to reduce the noise from the channel measurement and automatically adapts the threshold for each rate.

	\item
	\textit{ESNR}~\cite{esnr} is also an SNR-based algorithm. Different from CHARM, it takes frequency-selective fading into consideration. 
	%It predict the highest transmission rate by using the recently measured CSI. 
	The receiver measures CSI on the received packets and returns the optimal rate with the ACK. 

	\item
	\textit{OPT and Previous-OPT algorithms}~\cite{esnr}. 
	OPT algorithm is an omniscient algorithm that adds upper bounds for all the schemes. 
	OPT algorithm knows the highest transmission rate for the next frame that can be successfully sent.
	The Previous-OPT algorithm is also an ideal algorithm. It knows the optimal rate for the received frame and uses it for the next transmission. % esnr, rate selection algorithm
	The gap between the Previous-OPT and other algorithms is that Previous-OPT uses the real channel while other algorithms use the estimated channel.

\end{itemize}

% mobirate, overall performance 
\subsection{Overall Performance} 

% mobirate, overall performance; softrate, slow fading mobile channels
To begin with, we investigate the overall performance of StateRate in a fixed environment. 
First, we study the prediction accuracy when the UAV flies randomly.
Fig.~\ref{fig:overall_accuracy} shows the comparison of the transmission rate picked by various algorithms with the optimal rate for the next frame.
We observe that StateRate predicts the optimal rate with an accuracy of over 83.86\%, which is close to the Previous-OPT algorithm (84.97\%).

To better understand the overall performance of these algorithms, we evaluate the throughput of transmission using the same trace.
The results are shown in Fig.~\ref{fig:overall_throughput}.
As expected, the throughput of StateRate is $1.12\times$, $1.32\times$, $1.22\times$ of ESNR, SampleRate, and CHARM, respectively.
Another interesting point to note is the throughput of all the algorithms is highly related to the prediction accuracy, indicating that the prediction accuracy of the optimal rate can be used as the network output to improve the throughput.

 % pensieve, 5.2
\subsection{Impact of Velocities}

We further compare StateRate with the above algorithms in different velocities of the UAV.
In each experiment, we control the velocity of the UAV to fly within an appropriate range and collect the traces. 
StateRate is trained to predict the optimal rate using the entire training set collected from the same environment.

Fig.~\ref{fig:states_evaluation} provides detailed results of multiple rate adaptation algorithms.
There are some key observations from these results. 
First, it can be seen that StateRate exceeds the performance of the existing rate adaptation algorithm in each scenario.
Specifically, StateRate outperforms Previous-OPT because Previous-OPT has no knowledge about the future channel.
ESNR performs worse than Previous-OPT due to the inaccurate channel measurement.
% \textbf{Specifically, StateRate outperforms Previous-OPT because Previous-OPT has no knowledge about the future channel. ESNR performs worse than Previous-OPT due to the inaccurate channel measurement. The impact of the inaccurate channel also induces the gap between the ESNR and CHARM in the low-velocity scenario (0-2~$m/s$). However, ESNR is more suitable in a more dynamic environment because it predicts the bit rate based on the previous packet.}

Second, we observe that conventional rate adaptation algorithms struggle to adapt to different velocities.
The reason is that these algorithms employ the same parameters to predict MCS, even though different flight states require different strategies. 
For example, at a high-velocity scenario, the prediction needs to avoid using the long time window of the channel. 
However, to obtain the rate for the low-velocity scenario, a short time window is necessary.
The threshold for changing the transmission rate may also be different in different velocities.
StateRate can automatically learn these strategies.
Thus, when the velocity is 2-6~$m/s$, StateRate still improves the throughput by 53.01\% compared with ESNR.

The comparison between StateRate and OPT further illustrates the stability of StateRate. Stability refers to the closeness of StateRate's throughput to the upper bound. In detail, the prediction accuracies of StateRate across different velocities are 88.45\%, 82.45\%, and 82.7\%, respectively. Accordingly, the throughput of StateRate achieves 92.75\%, 92.07\%, and 88.61\% of the OPT algorithm. The throughput of StateRate is larger than 88\% of that of OPT despite that the throughput of OPT changing drastically.

\subsection{Evaluation for Different Trajectories} % esnr, rate adaptation results

\begin{figure}[t]
	\centering
	\includegraphics[width=0.5\textwidth]{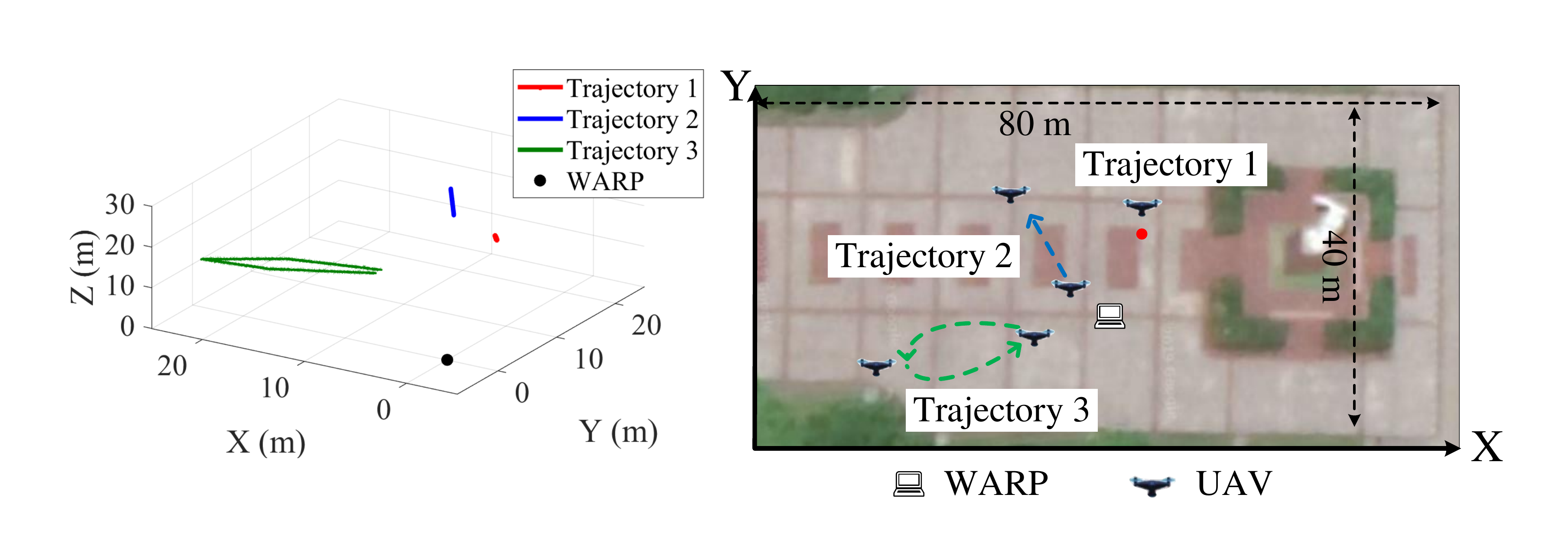}
	\caption{Three Trajectories.}
	% \vspace{-0.1cm}
	\label{fig:trajectories}
\end{figure}

\begin{figure*}
	\centering
	\subfigure[\scriptsize Trajectory 1]
	{
		\label{fig:hovering}
		\includegraphics[width=0.32\textwidth]{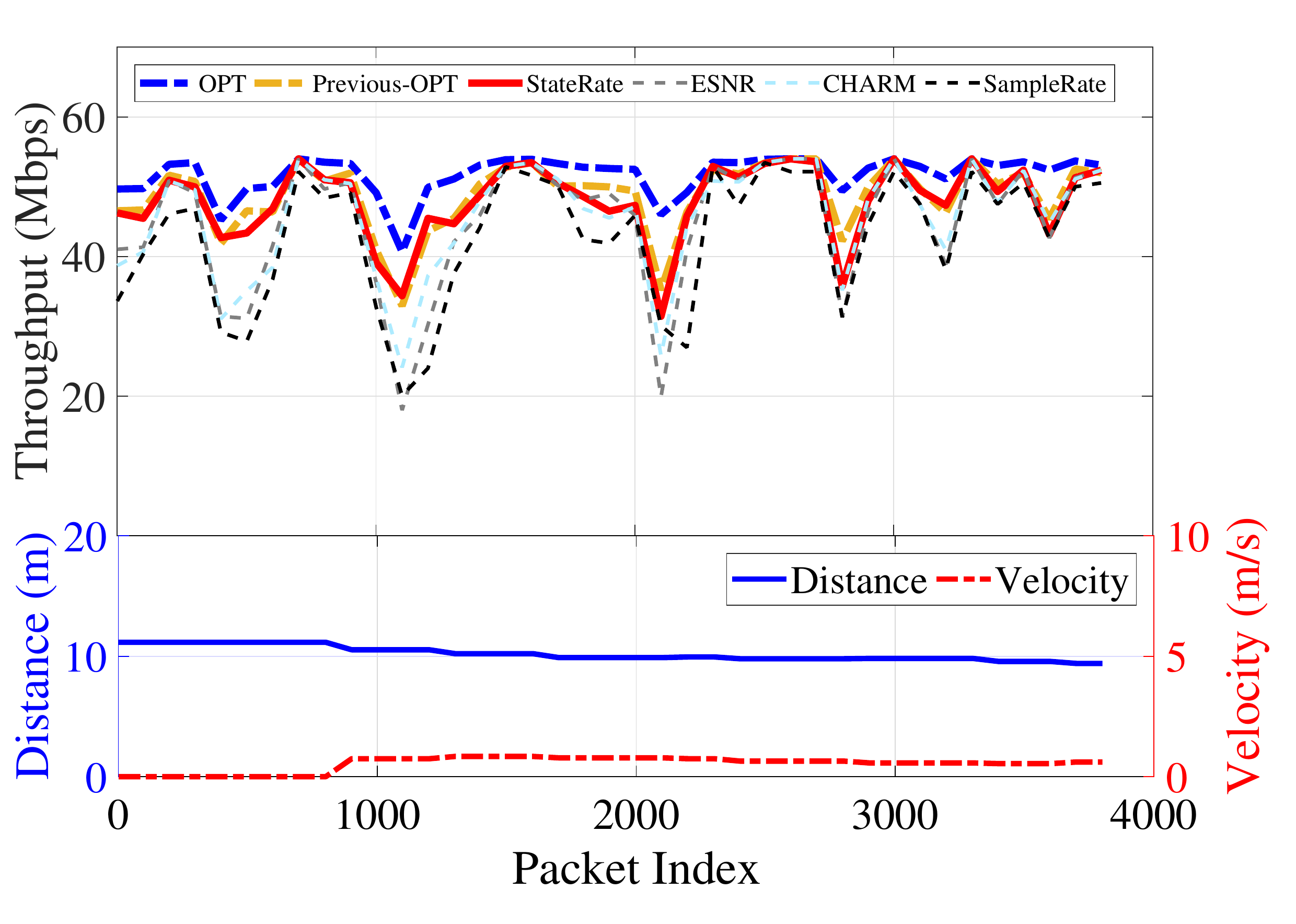}
	}
	\hspace{-0.4cm}
	\subfigure[\scriptsize Trajectory 2]
	{
		\label{fig:medium_velocity}
		\includegraphics[width=0.32\textwidth]{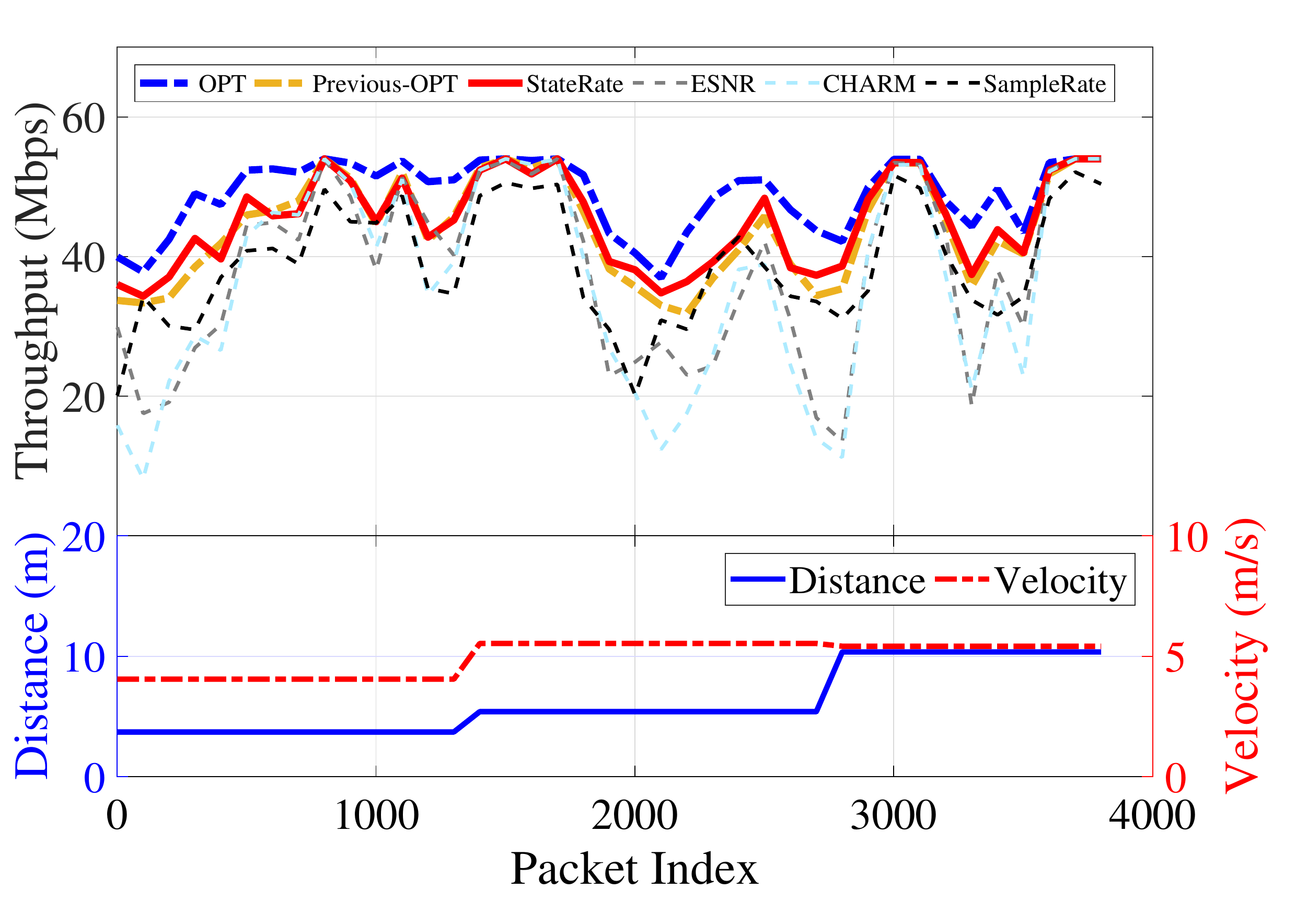}
	}
	\hspace{-0.4cm}
	\subfigure[\scriptsize Trajectory 3]
	{
		\label{fig:high_velocity}
		\includegraphics[width=0.32\textwidth]{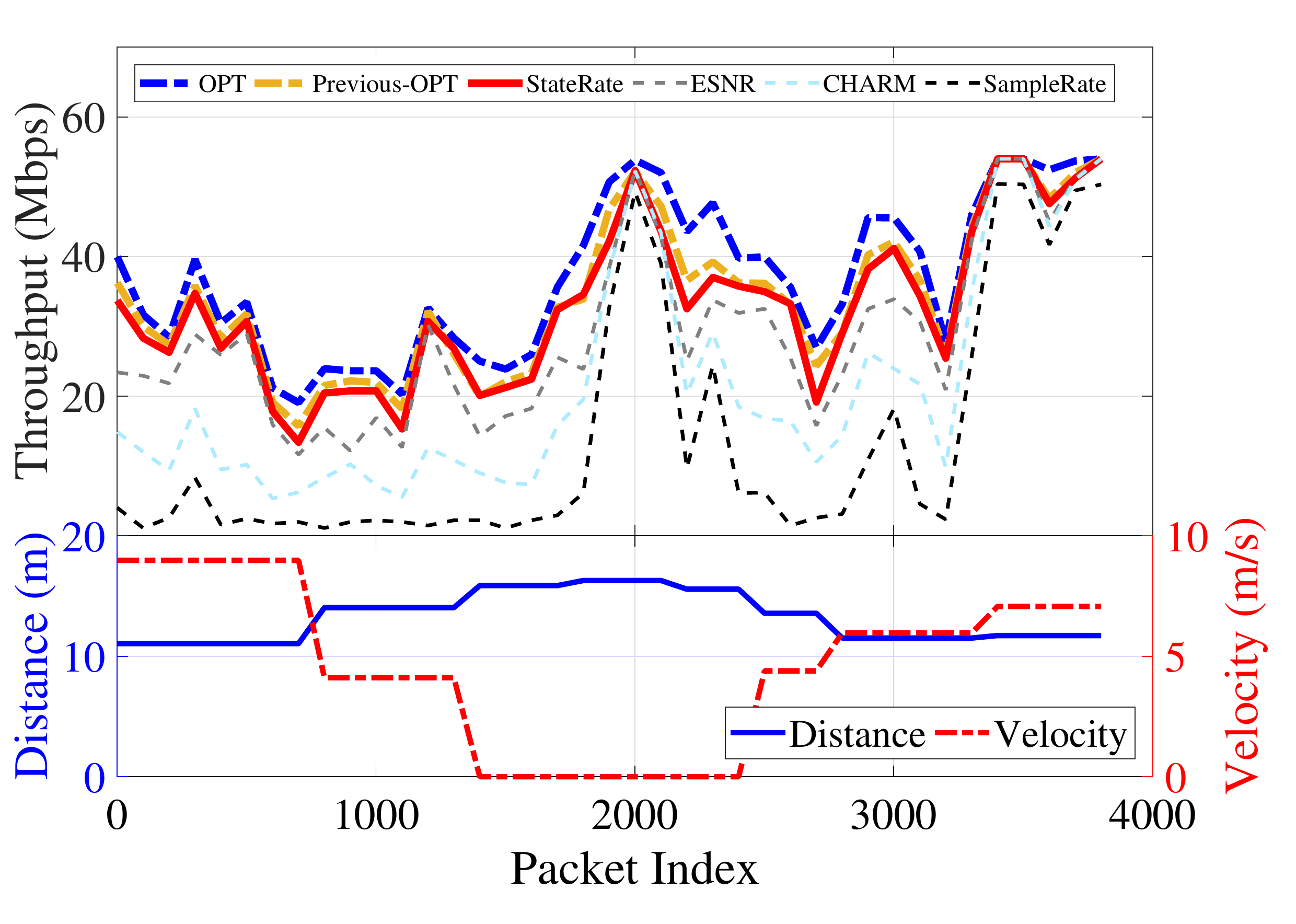}
	}
	\caption{Throughput under different trajectories.}
	\vspace{-0.15cm}
	\label{fig:traces_results}
\end{figure*}

The above experiments show that StateRate works well in different velocities.
In order to further evaluate the adaptability of our algorithm when flight states are dynamically switched, we examine the performance of StateRate under different trajectories. 
The trajectories in the environment are shown in Fig.~\ref{fig:trajectories}. The heights for three trajectories are fixed as 25~$m$, 30~$m$, 20~$m$, respectively, In addition, a part of the real-time flight states in these trajectories are shown at the bottom of Fig.~\ref{fig:traces_results}.
Three types of trajectories are evaluated in our experiments:

\begin{itemize}
	\item \textbf{Trajectory 1:} the UAV hovers at a horizontal distance of 10~$m$, which is a reasonable distance for practical scenarios, such as the follow-me mode in commercial UAVs~\cite{followme_cots}.
	
	\item \textbf{Trajectory 2:} the UAV flies from near to far at a constant velocity. The flight velocity is approximately 5~$m/s$.
	
	\item \textbf{Trajectory 3:} the UAV flies back and forth at a variable velocity. 
	
\end{itemize}

From these figures, we have the following observations: 

% esnr, siso performance; mobirate, overall performance 
\begin{itemize}
	\item[1)] Fig.~\ref{fig:hovering} illustrates the results when the UAV is hovering. 
	Note that it is difficult for a commercial UAV to maintain absolute hovering due to the wind and other environmental factors. The control system constantly compensates for the drift. In this case, all the algorithms have better performance under this condition compared with other trajectories.
	While it is hard to separate the lines, StateRate slightly outperforms CHARM which outperforms ESNR and SampleRate in most of the time. 
	
	\item[2)] The results of Trajectory 2 are shown in Fig.~\ref{fig:medium_velocity}.
	The throughput of all these algorithms including OPT and Previous-OPT is negatively impacted by mobility. 
	The three conventional algorithms do not adapt well in part of the time period under this condition. 
	However, StateRate is still comparable to Previous-OPT and outperforms other existing algorithms.
	
	\item[3)] Fig.~\ref{fig:high_velocity} shows the results of the UAV in a more dynamic scenario compared with the previous trajectories, where both velocity and distance change in this case. 
	As expected, the performance of StateRate is also close to the OPT and Previous-OPT at all times. However, the performance of other algorithms is also negatively impacted by dynamic velocity and distance. For example, the throughput of SampleRate is less than 10~Mbps when the UAV is moving. The gain of StateRate comes from the hints of the flight state by combining on-board sensors.
	% In addition, it can also be seen that the throughput of all algorithms significantly increases when the UAV is hovering at a distance of nearly $15~m$. 

	% \item[4)] Finally, the traditional algorithms are very unstable under the three trajectories, especially when the environment changes dramatically.
	% However, StateRate still achieves good performance all the time. 
	% It shows that the performance of StateRate is better than conventional algorithms not only in different states but also when the state changes dramatically.

\end{itemize}

\subsection{Evaluation for Deep Learning Framework}

In this section, we compare StateRate with sensor-hint rate adaptation algorithms to evaluate the advantage of employing a deep learning framework in StateRate. 
Previous work uses PHY hints~\cite{phy-hints} to improve the performance of rate adaptation. 
The problem of these algorithms is that they can only distinguish between motion and static scenarios.
Accurate velocity and distance are not used to optimize these algorithms. 
Therefore, we cannot implement the existing algorithms as the baseline. 
We design two simple sensor-augmented algorithms named Dynamic SampleRate and Dynamic ESNR, which are derived from SampleRate and ESNR, respectively.
The Dynamic SampleRate algorithm changes the size of the sliding window with the velocity and distance. 
The Dynamic ESNR algorithm adjusts the threshold based on the velocity which is obtained from sensors. 
In contrast, the network used in StateRate is trained using the training set, and the parameters are not adjusted after the training is completed.
We also exhibit the results of SampleRate and ESNR as the algorithms without the help of sensors.

Fig.~\ref{fig:throughput_sensorhints} shows the results of the above algorithms under different flight states. 
We notice that when the UAV hovers at a close distance, all algorithms achieve similar performance. 
In contrast, when the flight states change, the Dynamic ESNR and Dynamic SampleRate outperform ESNR and SampleRate, respectively. The gain of the two sensor-augmented algorithms comes from the change of the parameters according to the flight states. Another interesting observation is that the sensor-augmented algorithms are still worse than StateRate. This illustrates the difficulty of manually adjusting the parameters and the necessity of deploying a deep learning framework.
% Thus, we turn to propose an LSTM-based framework to take the temporal characteristics of the channel and automatically determine the parameters.}
% StateRate works well because the deep learning framework learns automatically the direction and the orientation of the UAV.

\subsection{Impact of Environments} % pensieve, generalization

In the above experiments, StateRate is trained with the traces collected in the same environments. 
In this experiment, the performance of StateRate in multiple environments is evaluated. 
The network is trained offline using the traces from the playground and then trained online with the traces from three other environments.
% The sites are shown in Fig.~\ref{fig:sites}.

To evaluate how StateRate generalizes to new environments, we conduct three experiments. 
First, we evaluate the accuracy of the evaluation network. The compared schemes are named SNR-evaluation and ESNR-evaluation, respectively. 
The SNR-evaluation scheme uses a well-known relationship between SNR and packet reception to verify whether the prediction is correct. 
The performance of this method is mainly affected by frequency-selective fading. 
The ESNR-evaluation scheme uses a similar setup to the SNR-evaluation scheme, except it uses ESNR~\cite{esnr} instead of SNR.
Second, we also compare StateRate using online learning with other rate adaptation algorithms. 
Third, we study the advantages of the online learning optimization and the pre-trained model.

\begin{figure*}
	\begin{minipage}[b]{0.45\textwidth}
		\centering
		\includegraphics[width=1\textwidth]{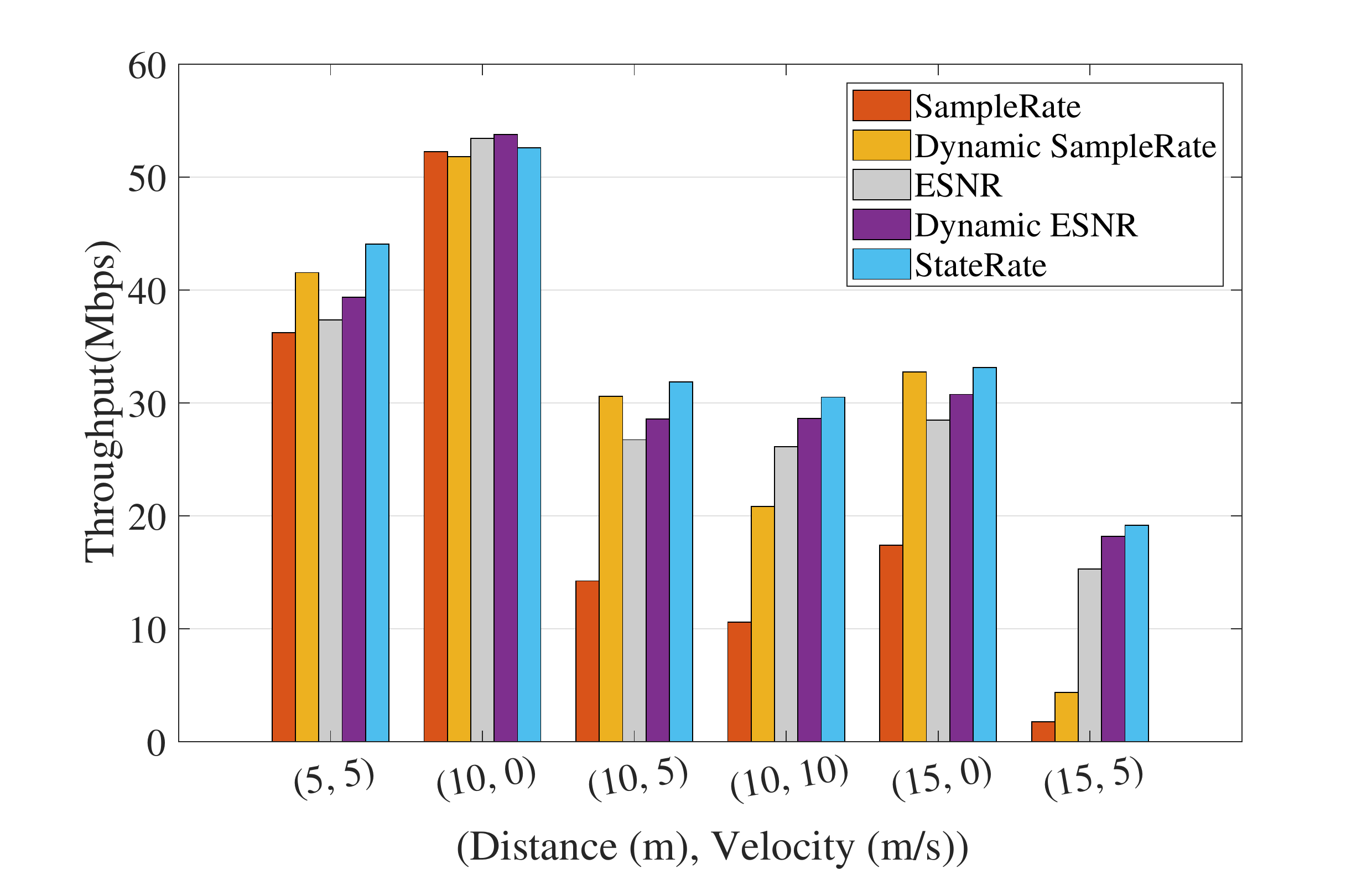}
		\caption{Compared with the sensor-augmented algorithm.}
		\label{fig:throughput_sensorhints}
	\end{minipage}
	\hspace{0.9cm}
	\begin{minipage}[b]{0.45\textwidth}
		\centering
		\includegraphics[width=1\textwidth]{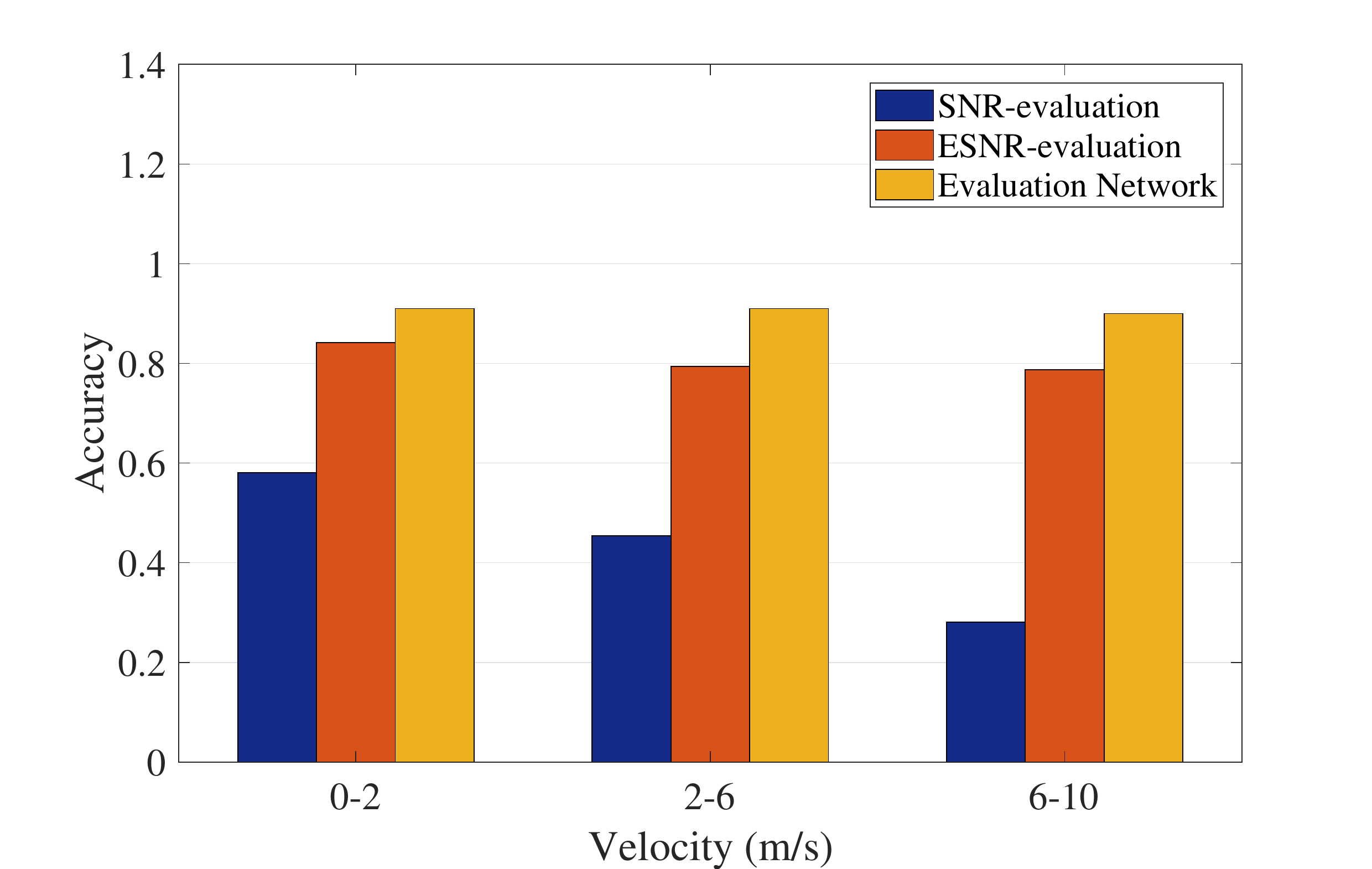}
		\caption{The accuracy of the evaluation network.}
		\label{fig:estimation_accuracy}
	\end{minipage}

	\subfigure[\scriptsize Compared with conventional methods]
	{
		\hspace{-0.2cm}
		\label{fig:throughput_stateoftheart}
		\includegraphics[width=0.45\textwidth]{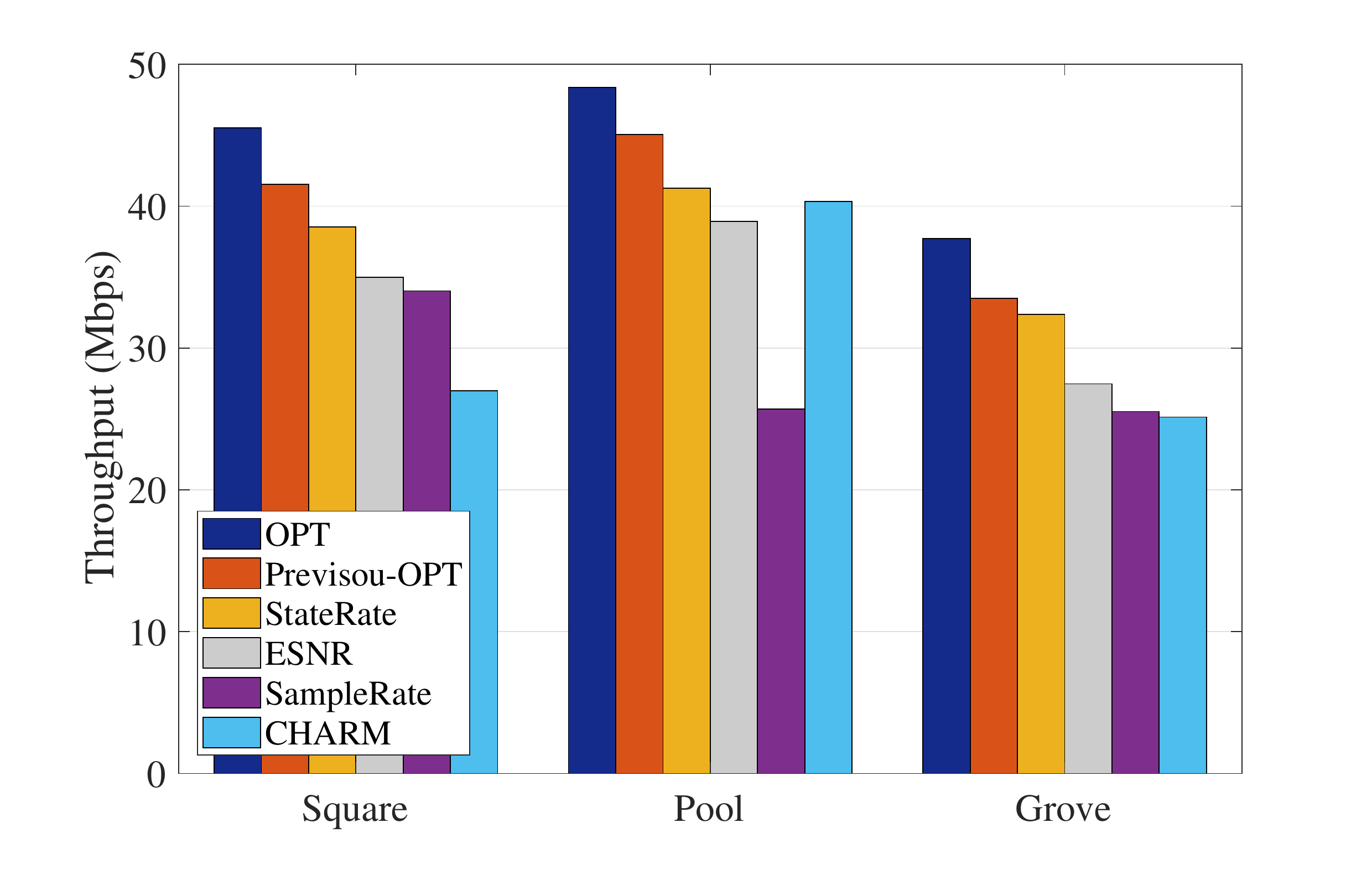}
	}
	\hspace{0.6cm}
	\subfigure[\scriptsize Compared with the retrained model and the model w/o online learning]
	{
		\label{fig:throughput_places}
		\includegraphics[width=0.45\textwidth]{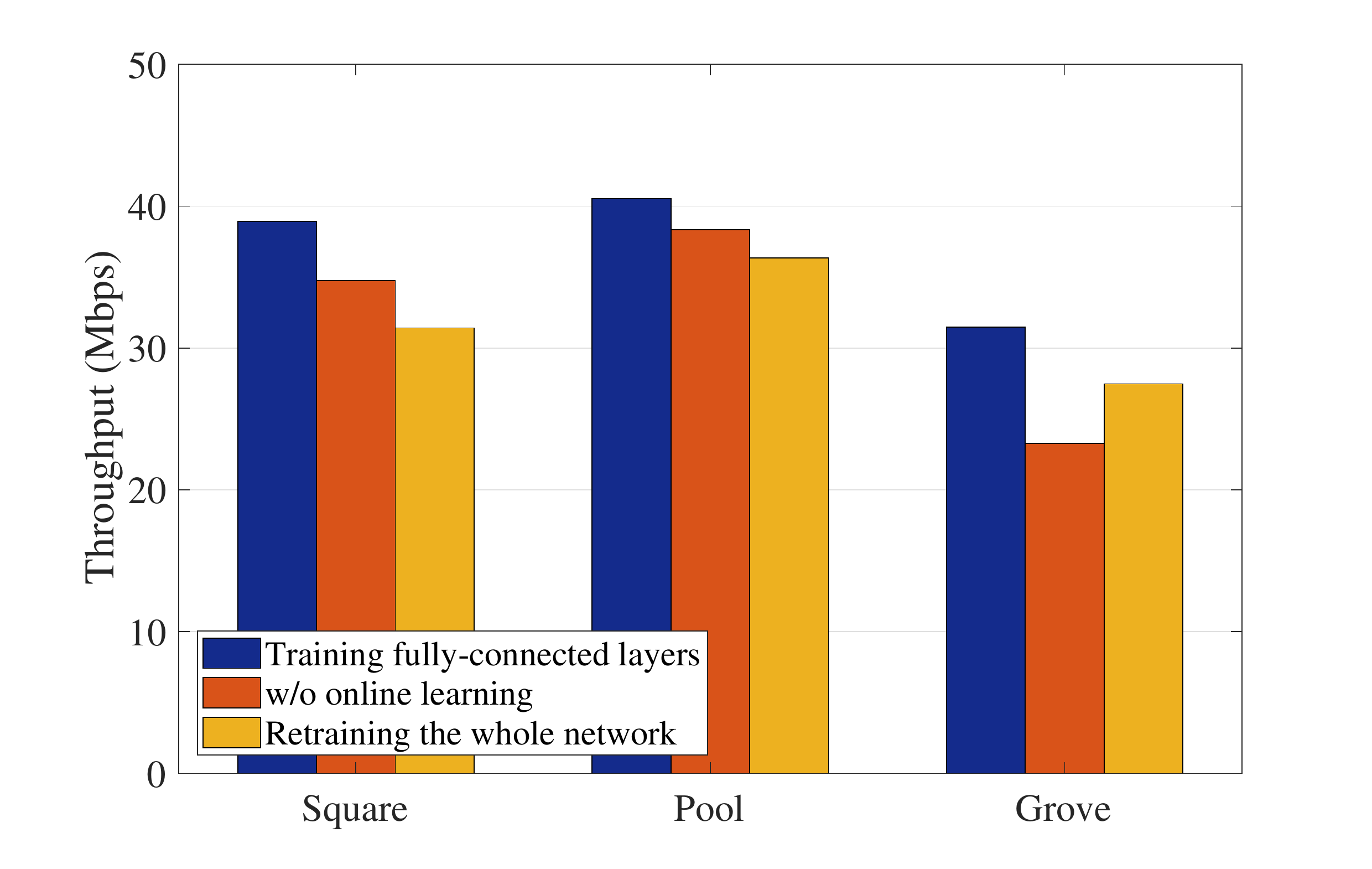}
	}
	\caption{Prediction performance under different environments.}
	\label{fig:environments}
\end{figure*}

We first investigate how the evaluation network performs and the results are shown in Fig.~\ref{fig:estimation_accuracy}. 
The average accuracy across different velocities is 90.67\%, which is adequate for online training. 
ESNR-evaluation also performs stably because the relationship between ESNR and the packet reception rate (PRR) is not impacted by the flight states and the environments.
The results finally demonstrate that the SNR-evaluation works more poorly as the velocity increases because profiles of frequency-selective fading vary significantly in mobility scenarios. 
% demonstrate, exhibit, depict

From Fig.~\ref{fig:throughput_stateoftheart}, StateRate also outperforms other conventional algorithms in different environments. 
In these three environments, StateRate can achieve an average of 85.2\% of the throughput compared with the OPT algorithm.
StateRate is also very stable with little fluctuation in different environments. 
In contrast, other algorithms seriously compromise throughput. 
The main reason is the conventional algorithms in different environments needing to change strategies due to different path losses and multipath components. 

Fig.~\ref{fig:throughput_places} illustrates the results of two disparate training methods to show the advantage of the pre-trained model. 
We observe that retraining the whole network is not the most suitable method because we collect a small amount of data from the new environment to train the model. 
However, we also observe that StateRate without online training performs poorly when the dataset of the pre-trained model and the online training model are significantly different. 
In specific, StateRate without online learning works even worse than the algorithm using the retrained model in the grove. 

% \vspace{0.2cm}
\section{Related Work}
\label{sec:relate}
%FARA, Softrate, ESNR, TiM, CCSI, strider, etc

Rate adaptation algorithms have been widely used in various communication technologies~\cite{noma}.
Most rate adaptation algorithms rely on one of two metrics: packet loss or channel state measurements. 

\textbf{Loss-based algorithm.}
Many loss-based algorithms have been used in lots of commercial devices~\cite{minstrel, ath9k}.
SampleRate~\cite{samplerate} is a famous solution in this category, and it works well in static environments. 
MiRA~\cite{mira} is the first work which considers the different loss pattern between single-stream and double-stream modes.
HA-RRAA~\cite{ha-rraa} uses an adaptive frame window and prevents collision losses by designing an RTS filter.
% The above algorithms are unable to respond promptly to motion-based channel changes.

In recent years, sensor fusion techniques have been adopted in the wireless localization system~\cite{BSDP, crowdsourcing_fusion}. These techniques utilize several types of sensor data to improve wireless localization performance.
Derived by the multisource fusion in the wireless localization system, several works focus on optimizing indoor mobile communications~\cite{sensor-hints, mobirate} and outdoor vehicular communications~\cite{lora,joint,conditional} with sensors.
For example, the mobility hints are proposed in ~\cite{sensor-hints} to switch between SampleRate and RapidSample.
In the outdoor vehicular scenarios, CARS~\cite{cars} uses GPS information to address the underutilization of link capacity due to the link fading and infrequent transmission. 
However, these algorithms are designed to have advantages in the vehicle environment and are not optimized for UAV flight states.
% Similar to our design, SA-ABR~\cite{saabr} investigates the impact of flight states to the throughput, and design a reinforcement learning framework to optimize the video quality.

\textbf{SNR-based algorithm.}
The SNR-based approaches assume that the optimal rate can be chosen based on SNR-PRR relation. 
CHARM~\cite{charm} uses signal strength to choose the optimal rate. 
It is difficult for CHARM to accurately find the SNR-PRR relationship in the OFDM system due to the frequency-selective fading. 
To solve this problem, FARA~\cite{fara} computes SNR for each subcarrier and picks a rate that matches the SNR. 
ESNR~\cite{esnr} introduces a new metric named ESNR to combat the frequency-selective fading. 
In addition, Smart Pilot~\cite{smartpilot} calibrates CSI by combining packet headers from the upper layer.
Softrate~\cite{softrate} and AccuRate~\cite{accurate} bring SoftPHY hints to estimate BER for each packet even when the packet has no errors. 
These approaches add an additional module at receivers to opt out packet loss caused by collision.
Differently, Strider~\cite{strider} designs a collision-resilient rate adaptation by extending rateless codes without such an extra module.
TiM~\cite{tim} focuses on inserting modulation schemes between two existing schemes by adding time domain diversity in modulation.
A mobility-aware rate adaptation algorithm~\cite{phy-hints} is proposed to use PHY layer hints extracted from CSI and time of flight (ToF) to improve network performance. 
SmartLA\cite{smartla} is a reinforcement learning mechanism for dynamic selection of link parameters, while it does not combine sensors.

% Different from these approaches, StateRate is the first work for the UAV rate adaptation algorithm by combining the on-board sensors with PHY hints under dynamic conditions.
Although some studies~\cite{saabr, skyeyes} optimize video quality for UAVs with the help of sensors, StateRate is the first work for the UAV rate adaptation algorithm by fusing the on-board sensors with PHY hints under dynamic conditions.
We design a deep learning framework to dramatically predict transmission rate using both the channel states and the flight states.

\section{Discussion}
\label{sec:discussion}

We discuss some practical considerations of StateRate, including compatibility, training cost, computational overhead and use cases.

\textbf{Compatibility.}
StateRate uses information that can be directly obtained from commercial devices. It can be easily applied to existing commercial wireless network cards and fully complies with IEEE 802.11 protocols. Additionally, as rate adaptation is a manufacturer-dependent algorithm whose strategies are not specified in the standards. Thus, our scheme can also work with other protocols and future protocols.

\textbf{Training cost.} 
We optimize the rate by training a deep learning network. The algorithm needs a large amount of training data with enough labels. It will inevitably spend lots of time to train the offline model, which leads to expensive training cost. Fortunately, training cost is unnecessary for commercial UAVs as we can train the whole network offline. The offline training process does not add burdens to UAVs. In addition, our online training only needs to update a part of the parameters, which reduces the online training cost.

\textbf{Computational overhead.} 
The proposed algorithm is more complex than conventional rate adaptation algorithms due to the complexity of the deep learning structure, which increases the computational overhead. Existing UAVs require vision sensors for navigation and object recognition, which are also based on the deep learning framework. Compared with these algorithms, our learning-based rate adaptation algorithm incurs much lower computational overhead due to its simpler structure. Thus, the computational overhead is totally affordable for UAVs.

\textbf{Use cases.}
We believe that the proposed algorithm can improve the reliability and reduce the delay for future air-to-ground transmission. As an example, in the SAR missions where UAVs are required to send rescue information to the server within a short delay, StateRate can reduce transmission delays, saving lots of time for recovering victims. Another example is that UAVs equipped with high-definition cameras and microphones can fly above a crowded stadium, filming videos and streaming them to users on the ground. StateRate can also guarantee the stability of video transmission.

\section{Conclusion and Future Work}
\label{sec:conclusion}

We present StateRate, a state-optimized rate adaptation algorithm for air-to-ground links.
Unlike conventional SNR-based algorithms, StateRate predicts the optimal rate with the assistance of the sensors on the UAV. 
StateRate employs a deep learning architecture that learns to choose the optimal MCS when the state of the UAV changes dynamically. 
Moreover, we develop an online learning algorithm with a pre-trained model to adapt to different environments. 
We verify the performance of StateRate under various flight states and environments.
The results reveal that StateRate improves the throughput by up to 53\% compared with the best-known rate adaptation algorithm when the velocity is 2-6~$m/s$.

The proposed scheme performance can be further improved by employing Rician model as an orthogonal input. In specific, the statistical information extracted from this model can be leveraged to improve the performance of the rate adaptation. However, incorporating the Rician model into rate adaptation is still challenging, since it is difficult to estimate the parameters in an unknown environment. Thus, it requires brand new observations to incorporate the Rician model into our work and we intend to leave it for future consideration.

In addition to the advantages of the Rician model, rate and frame aggregation length adaptation for multi-antenna UAVs is a worthwhile direction for future investigation. IEEE 802.11ac standard supports up to eight antennas, which can significantly increase the throughput of UAV communication. Furthermore, frame aggregation can also be used to improve the transmission quality in the case of video streaming or bulk data downloading. Combining multiple antennas and frame aggregation with rate adaptation requires a strategy that is more complex than just considering modulation and coding.

% \balance

\bibliographystyle{IEEEtran}
\bibliography{IEEEabrv,./RateAdaptation}

\vspace{-0.5cm}

\begin{IEEEbiography}
	[{\includegraphics[width=1in,height=1.25in,clip,keepaspectratio]{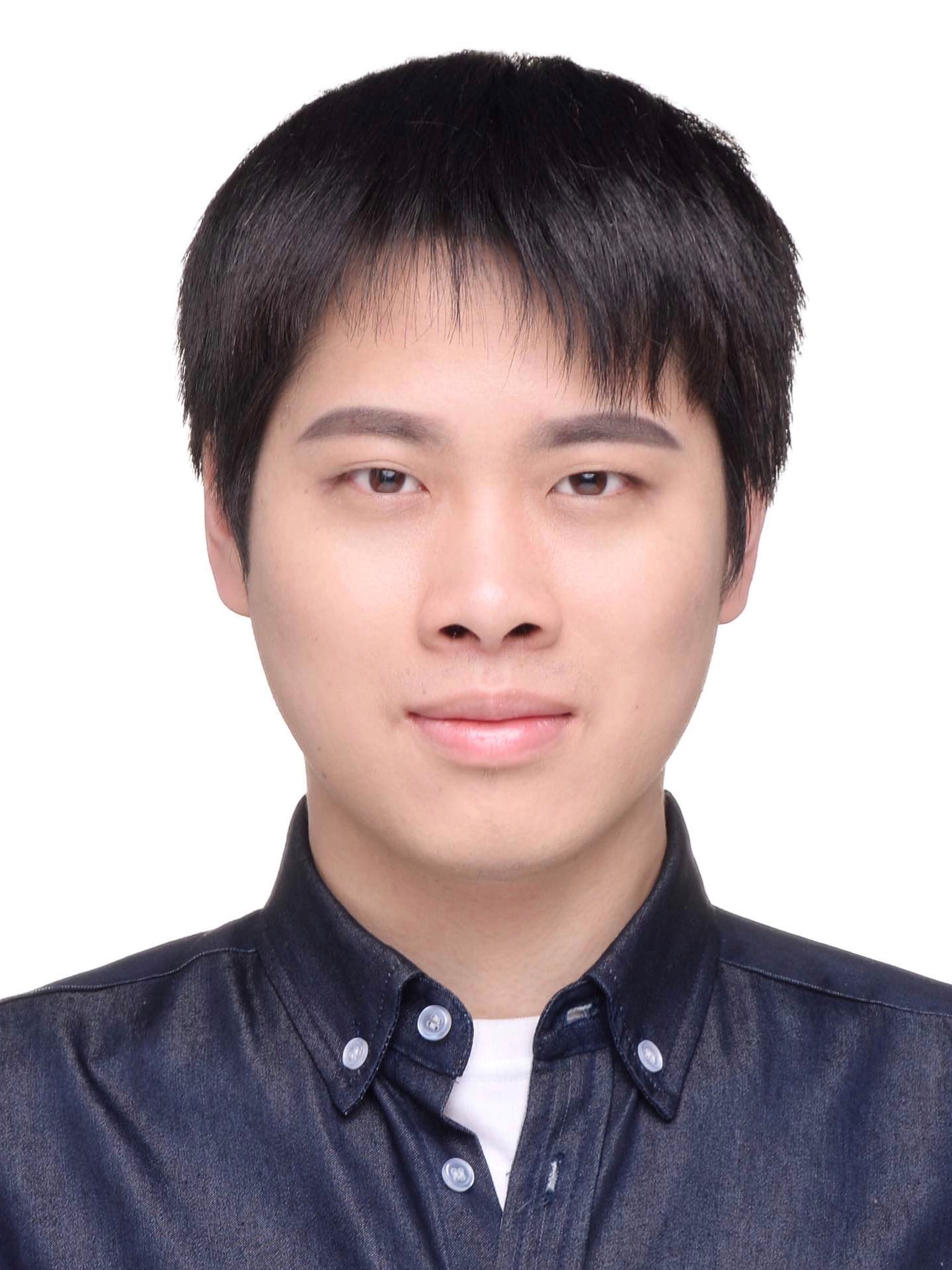}}]
	{Shiyue He} is currently pursuing his Ph.D. degree at School of Electronics and Information Engineering, Huazhong University of Science and Technology, Hubei, China. Before that, he has received his Bachelors degree in information engineering from Wuhan University of Technology, Hubei, China, in June 2016. His research interests include UAV communications and beamforming in wireless networks.
\end{IEEEbiography}

\vspace{-1cm}

\begin{IEEEbiography}[{\includegraphics[width=1in,height=1.25in,clip,keepaspectratio]{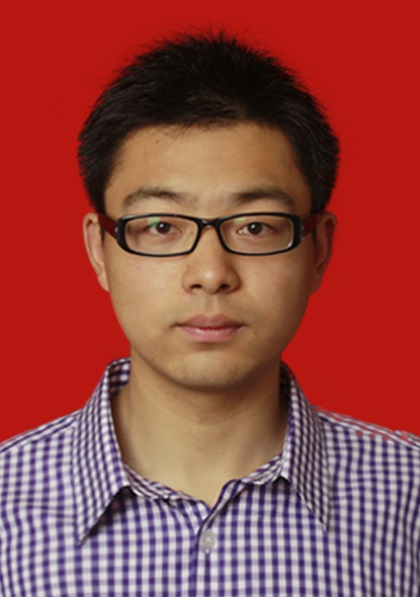}}]
	{Wei Wang (S'10-M'16)} received the Ph.D. degree from the Department of Computer Science and Engineering, The Hong Kong University of Science and Technology. He is currently a Professor with the School of Electronic Information and Communications, Huazhong University of Science and Technology. His research interests include PHY/MAC design and mobile computing in wireless systems. He served on TPC of INFOCOM and GBLOBECOM. He served as Editors for IJCS, China Communications, and Guest Editors for Wireless Communications and Mobile Computing and the IEEE COMSOC MMTC COMMUNICATIONS.
\end{IEEEbiography}

\vspace{-1cm}

\begin{IEEEbiography}
	[{\includegraphics[width=1in,height=1.25in,clip,keepaspectratio]{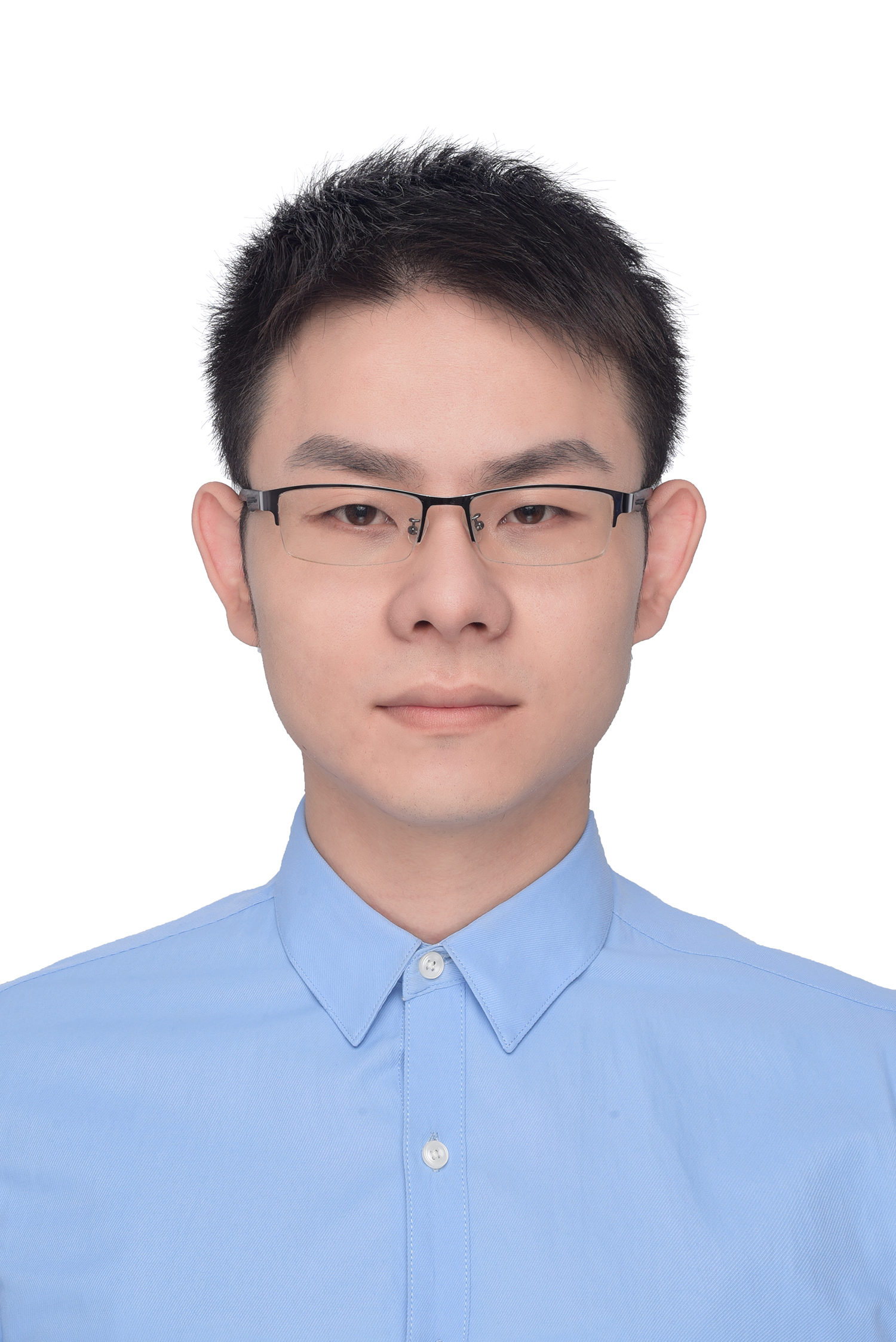}}]
	{Hang Yang} is currently pursuing his M.Phil. degree at School of Electronics and Information Engineering, Huazhong University of Science and Technology, Hubei, China. Before that, he received his Bachelors degree at there in June 2018. His research interests include channel coding and UAV intelligent communication in wireless network.
\end{IEEEbiography}

\vspace{-1cm}

\begin{IEEEbiography}[{\includegraphics[width=1in,height=1.25in,clip,keepaspectratio]{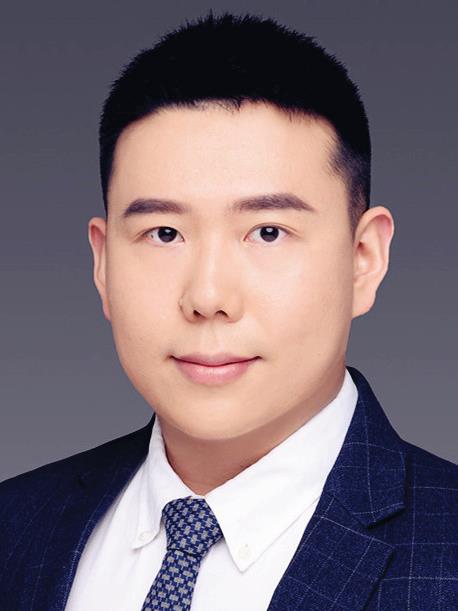}}]
	{Yang Cao (S'09-M'14)} received the B.S. and Ph.D. degrees in information and communication engineering from Huazhong University of Science and Technology, Wuhan, China, in 2009 and 2014. He is currently an Associate Professor in the School of Electronic Information and Communications, Huazhong University of Science and Technology, Wuhan, China. From 2011 to 2013, he worked in the School of Electrical, Computer, and Energy Engineering, Arizona State University, Tempe, AZ, USA, as a Visiting Scholar. His research interests include 5G cellular networks, Internet of Things, and future networks. He has coauthored 40 papers on refereed IEEE journals and conferences. He received CHINACOM Best Paper Award in 2010 and Microsoft Research Fellowship in 2011.
\end{IEEEbiography}

\vspace{-1cm}

\begin{IEEEbiography}[{\includegraphics[width=1in,height=1.25in,clip,keepaspectratio]{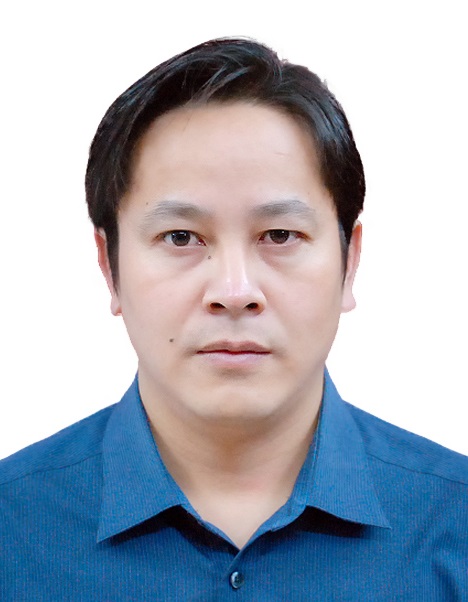}}]
	{Tao Jiang (M'06, SM'10,F'19)} is currently a Distinguished Professor in the Wuhan National Laboratory for Optoelectronics and School of Electronics Information and Communications, Huazhong University of Science and Technology, Wuhan, P. R. China. He received Ph.D. degree in information and communication engineering from Huazhong University of Science and Technology, Wuhan, P. R. China, in April 2004. From Aug. 2004 to Dec. 2007, he worked in some universities, such as Brunel University and University of Michigan-Dearborn, respectively. He has authored or co-authored more 300 technical papers in major journals and conferences and 9 books/chapters in the areas of communications and networks. He served or is serving as symposium technical program committee membership of some major IEEE conferences, including INFOCOM, GLOBECOM, and ICC, etc.. He was invited to serve as TPC Symposium Chair for the IEEE GLOBECOM 2013, IEEEE WCNC 2013 and ICCC 2013. He is served or serving as associate editor of some technical journals in communications, including in IEEE Network, IEEE Transactions on Signal Processing, IEEE Communications Surveys and Tutorials, IEEE Transactions on Vehicular Technology, IEEE Internet of Things Journal, and he is the associate editor-in-chief of China Communications, etc..
\end{IEEEbiography}

\vspace{-12cm}

\begin{IEEEbiography}[{\includegraphics[width=1in,height=1.25in,clip,keepaspectratio]{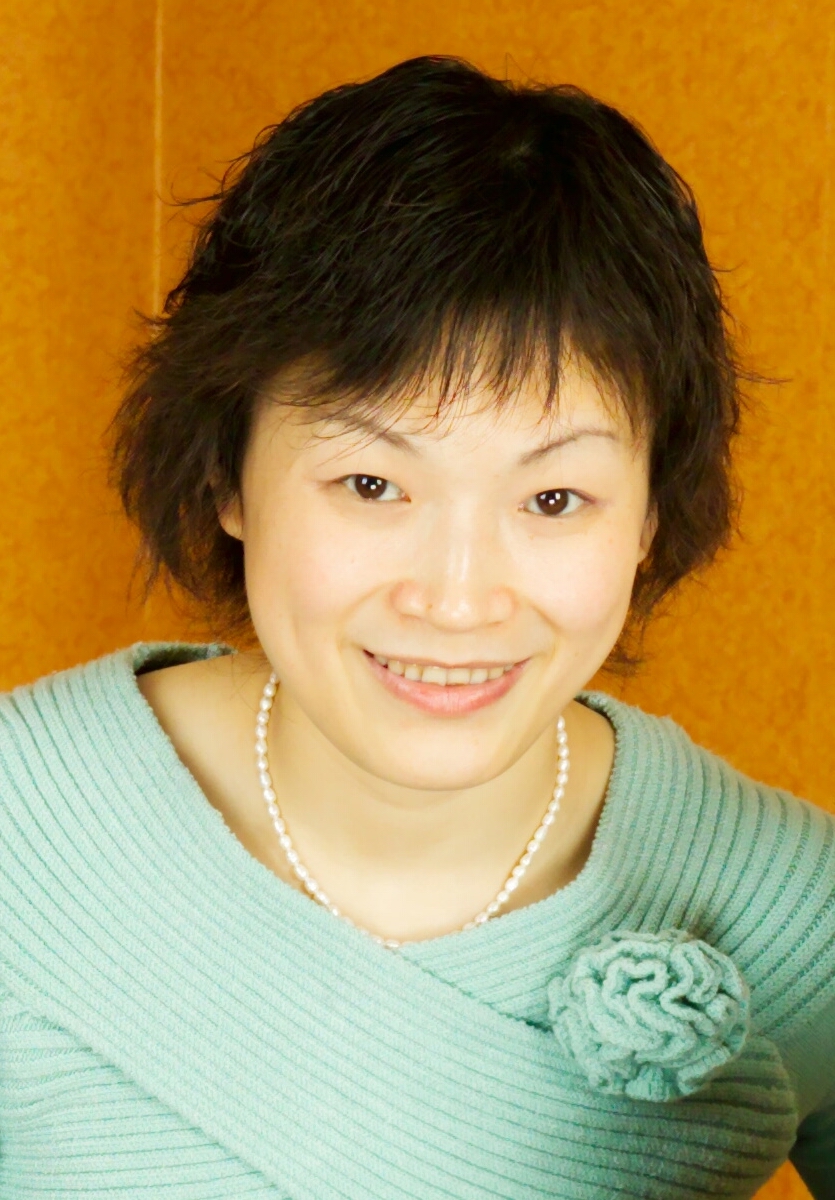}}]
	{Qian Zhang (M'00-SM'04-F'12)} joined Hong Kong University of Science and Technology in Sept. 2005 where she is a full Professor in the Department of Computer Science and Engineering. Before that, she was in Microsoft Research Asia, Beijing, from July 1999, where she was the research manager of the Wireless and Networking Group. She is a Fellow of IEEE for “contribution to the mobility and spectrum management of wireless networks and mobile communications”. Dr. Zhang received the B.S., M.S., and Ph.D. degrees from Wuhan University, China, in 1994, 1996, and 1999, respectively, all in computer science.
\end{IEEEbiography}

\end{document}